\documentclass[10pt,twocolumn,letterpaper]{article}

\usepackage[pagenumbers]{cvpr} 

\usepackage[ruled,vlined]{algorithm2e}
\usepackage{xcolor}
\usepackage{pifont}
\newcommand{\redcross}{\textcolor{red}{\textit{\ding{55}}}} 

\newcommand{\greencheck}{\textcolor{green}{\textit{\ding{52}}}}

\definecolor{cvprblue}{rgb}{0.21,0.49,0.74}
\usepackage[pagebackref,breaklinks,colorlinks,allcolors=cvprblue]{hyperref}

\def\paperID{2779} 
\def\confName{CVPR}
\def\confYear{2026}

\title{TagSplat: Topology-Aware Gaussian Splatting for Dynamic Mesh Modeling and Tracking}

\author{
\begin{tabular}{@{}c@{\hspace{20pt}}c@{\hspace{20pt}}c@{}}
\textbf{Hanzhi Guo} & \textbf{Dongdong Weng} & \textbf{Mo Su} \\
Beijing Institute of Technology & Beijing Institute of Technology & Soul Shell Technology Co., Ltd \\
Beijing & Beijing & Beijing \\
{\tt\small hanzhiguo@bit.edu.cn} & {\tt\small crgj@bit.edu.cn} & {\tt\small schumer425@126.com} \\[10pt]
\textbf{Yixiao Chen} & \textbf{Dongye Xiaonuo} & \textbf{Chenyu Xu} \\
Beijing Institute of Technology & Beijing Institute of Technology & Soul Shell Technology Co., Ltd \\
Beijing & Beijing & Beijing \\
{\tt\small yxchengeorge@163.com} & {\tt\small dyxn@bit.edu.cn} & 
\end{tabular}
}

\begin{document}
 
\twocolumn[{%
  \renewcommand\twocolumn[1][]{#1}
  \maketitle  
    \begin{center}
      \vspace{-0.8cm}
      \includegraphics[width=\textwidth]{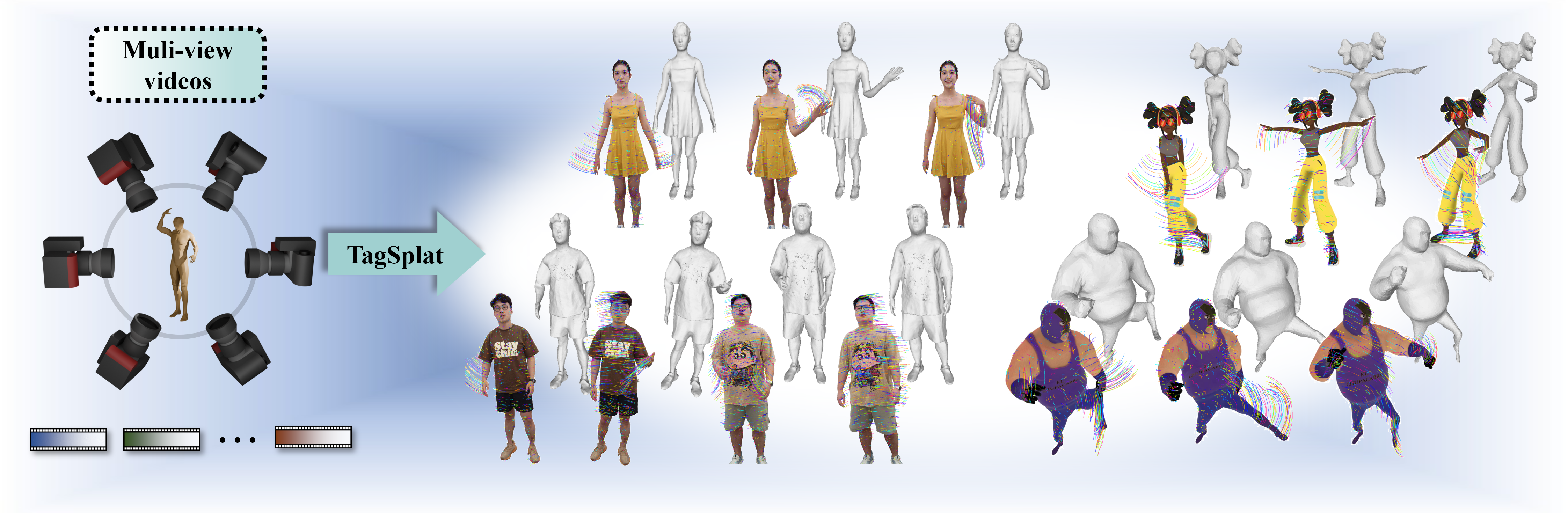}
      \vspace{-0.7cm}
      \captionof{figure}{Our method reconstructs dynamic humans from multi-view videos. Our topology-aware strategy produces topology-consistent Gaussian and mesh sequences, enabling accurate dynamic reconstruction and tracking.}
    \end{center}
}]

\begin{abstract}


Topology-consistent dynamic model sequences are essential for applications such as animation and model editing. However, existing 4D reconstruction methods face challenges in generating high-quality topology-consistent meshes. To address this, we propose a topology-aware dynamic reconstruction framework based on Gaussian Splatting. We introduce a Gaussian topological structure that explicitly encodes spatial connectivity. This structure enables topology-aware densification and pruning, preserving the manifold consistency of the Gaussian representation. Temporal regularization terms further ensure topological coherence over time, while differentiable mesh rasterization improves mesh quality. Experimental results demonstrate that our method reconstructs topology-consistent mesh sequences with significantly higher accuracy than existing approaches. Moreover, the resulting meshes enable precise 3D keypoint tracking. Project page: \url{https://haza628.github.io/tagSplat/}

\end{abstract}    
\section{Introduction}
\label{sec:intro}


The animation industry relies on a comprehensive mesh-centered toolchain for rendering, skinning, and editing. Efficiently reconstructing dynamic mesh sequences with consistent topology remains a key challenge bridging computer vision and graphics \cite{xue2023nsf, hanocka2020point2mesh, collet2015high, chen2023neusg, wang2021neus, guedon2024sugar}. Such topology-consistent meshes are crucial for downstream tasks. In traditional production pipelines, they are created manually or via optical-flow-based retopology of high-resolution per-frame reconstructions. Unlike costly manual pipelines, Gaussian Splatting provides an efficient explicit representation for high-quality 3D reconstruction\cite{kerbl20233d}. By explicitly modeling geometry and appearance with 3D Gaussians, it achieves fast reconstruction, high-fidelity rendering, and easy editability. Unlike implicit representations \cite{mildenhall2021nerf, park2019deepsdf, mescheder2019occupancy}, it preserves a point cloud structure, allowing seamless integration with standard geometric processing.

Recent studies have started exploring dynamic mesh generation from Gaussian Splatting. A 3D Gaussian representation can be treated as a point cloud. Meshes can then be reconstructed from this point cloud using methods such as Poisson reconstruction or depth-based techniques. Several works have attempted to reconstruct dynamic mesh sequences from Gaussian Splatting\cite{liu2024dynamic, zhang2024dynamic, zheng2025gaustar}. Reconstructed meshes are generated independently for each frame due to the lack of temporal consistency constraints. This independence leads to frame-wise variations in topology. As a result, downstream tasks such as skeleton binding and keypoint tracking remain fundamentally challenging. Therefore, how to generate topology-consistent dynamic mesh sequences remains an open problem.

\begin{figure}
    \centering
    \includegraphics[width=\linewidth]{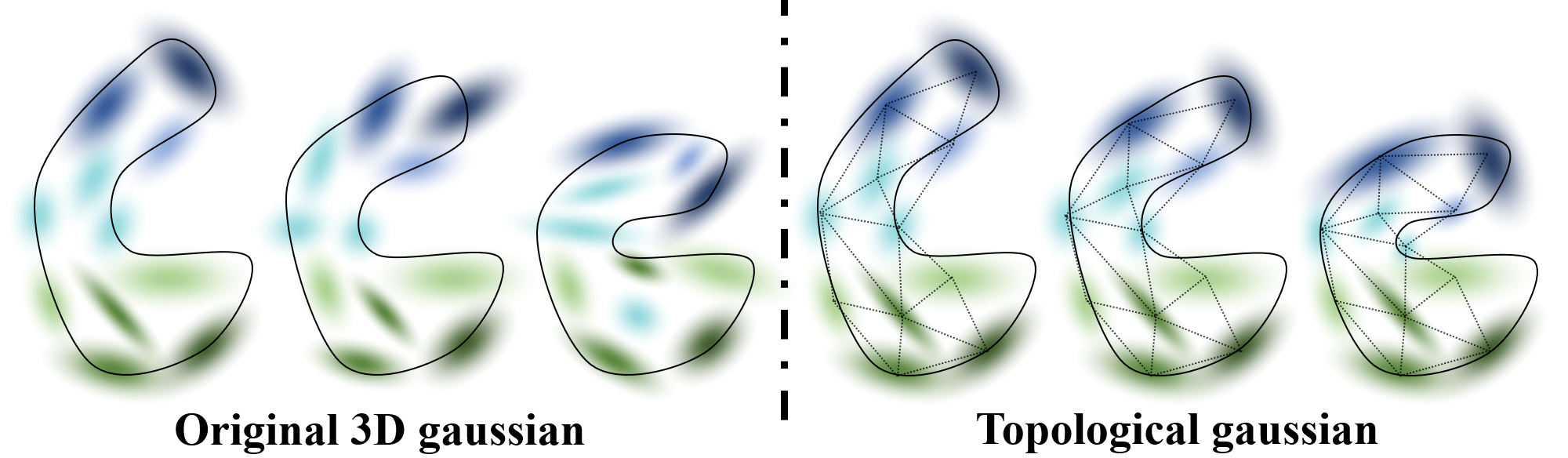}
    \vspace{-0.5cm}
    \caption{Illustration of the Gaussian topology structure. Each Gaussian primitive is connected via manifold edges. Compared with the original 3D Gaussians, our structure constrains the relative positions and rotations of neighboring Gaussians. This method enables training Gaussian and mesh models with consistent topology.}
    \vspace{-0.7cm}
    \label{fig:topo-gs}
\end{figure}

To address the challenge of reconstructing topology-consistent mesh sequences in Gaussian Splatting, we propose a Gaussian topological structure, as illustrated in \cref{fig:topo-gs}. Based on this structure, we propose a topology-aware 3D Gaussian densification and pruning process. The topology is automatically updated whenever Gaussian primitives are added or removed. This adaptive topology update ensures that the manifold structure is preserved during 3D Gaussian densification. The mesh vertices of each frame follow the positions of the corresponding 3D Gaussians, and combined with the Gaussian topology, the mesh model of the target object can be obtained. To ensure temporal coherence across dynamic sequences, we incorporate regularization terms that preserve the relative positions and rotations of each Gaussian primitive within its 1-ring neighborhood. We also employ a differentiable mesh rasterizer to improve geometric accuracy. This combination allows not only topology-consistent mesh reconstruction but also 3D keypoint tracking by preserving the relative positions of keypoints on the mesh surface over time. Our method is the first to maintain manifold topology through dynamic Gaussian updates, enabling the reconstruction of topology-consistent meshes and accurate tracking.

Our main contributions are summarized as follows:

\begin{itemize}
    \item We propose a Gaussian topology representation that bridges the gap between Gaussian Splatting and mesh models, enabling fast and consistent conversion.
    \item We propose an adaptive topology-based densification and pruning method that preserves the manifold topology during the Gaussian densification and pruning processes.
    \item We design temporal consistency constraints based on a Gaussian topology structure to ensure temporally smooth and topology-consistent Gaussian distributions in dynamic sequences.
    \item We build an end-to-end framework for dynamic reconstruction and 3D tracking, providing a practical and low-cost solution for animation and model editing tasks.
\end{itemize}

\section{Related works}
\label{sec:Related works}

\subsection{Neural 3D reconstruction and rendering}
Neural implicit representations refer to techniques that use neural networks to implicitly model 3D shapes and scene attributes. The most notable example is NeRF\cite{mildenhall2021nerf}, which optimizes voxel density and radiance from multi-view images to achieve high-fidelity novel view synthesis. Following NeRF, many works have focused on improving representation efficiency and detail quality. Instant-NGP\cite{muller2022instant} introduced hash-grid encoding to enable high-resolution NeRF training within minutes; Plenoxels\cite{fridovich2022plenoxels} directly optimize voxel attributes in a sparse voxel grid, avoiding the over-parameterization of neural networks.

To better adapt to complex geometry, NeuS\cite{wang2021neus} incorporated signed distance function (SDF) constraints and optimized surface normal consistency through differentiable rendering. VolSDF\cite{yariv2021volume} combined volumetric rendering with SDF to guide volume density concentration on differentiable surfaces. These methods achieve excellent results in static scene modeling but offer limited support for dynamic scenes or editable meshes.

Compared to NeRF, Gaussian Splatting provides an explicit and lightweight-optimizable 3D representation. Gaussian splatting\cite{kerbl20233d} systematically proposed 3D Gaussian splatting rendering, converting NeRF volumetric integration into differentiable rendering of continuous Gaussian distributions. This approach enables high-quality novel view synthesis within minutes and real-time performance on mobile devices.

The main advantage of Gaussian Splatting lies in explicitly modeling each Gaussian primitive’s position, covariance, and color attributes, which are easy to edit, merge, and densify. 
Numerous extensions of Gaussian Splatting have been explored\cite{huang20242d, dongye2024lodavatar, chen2025ps, gaussian_grouping, relightable}.
Dynamic scene reconstruction based on Gaussian Splatting has seen a surge of research efforts. Some studies focus on 4D reconstruction, recovering dynamic Gaussian representations of scenes from single-view or multi-view videos\cite{jiang2024hifi4g, luiten2024dynamic, jiang2025topology, jung2023deformable}. Others concentrate on human performance modeling, aiming to produce highly realistic and controllable human avatars\cite{gaussianheadavatar, hu2024gaussianavatar, li2024animatable, qian2024gaussianavatars, moreau2024human}. There are also approaches that introduce physical constraints into the Gaussian Splatting framework to learn physically consistent dynamic objects\cite{xie2024physgaussian, abou2024physically, choi2024phys3dgs}. In addition, some works address model compression to improve the efficiency and practicality of Gaussian Splatting in dynamic settings\cite{chen2025hac++, chen2024hac, chen2025pcgs, liu2024compgs, liu2025compgs++}.

\begin{figure*}
    \centering
    \vspace{-0.6cm}
    \includegraphics[width=\linewidth]{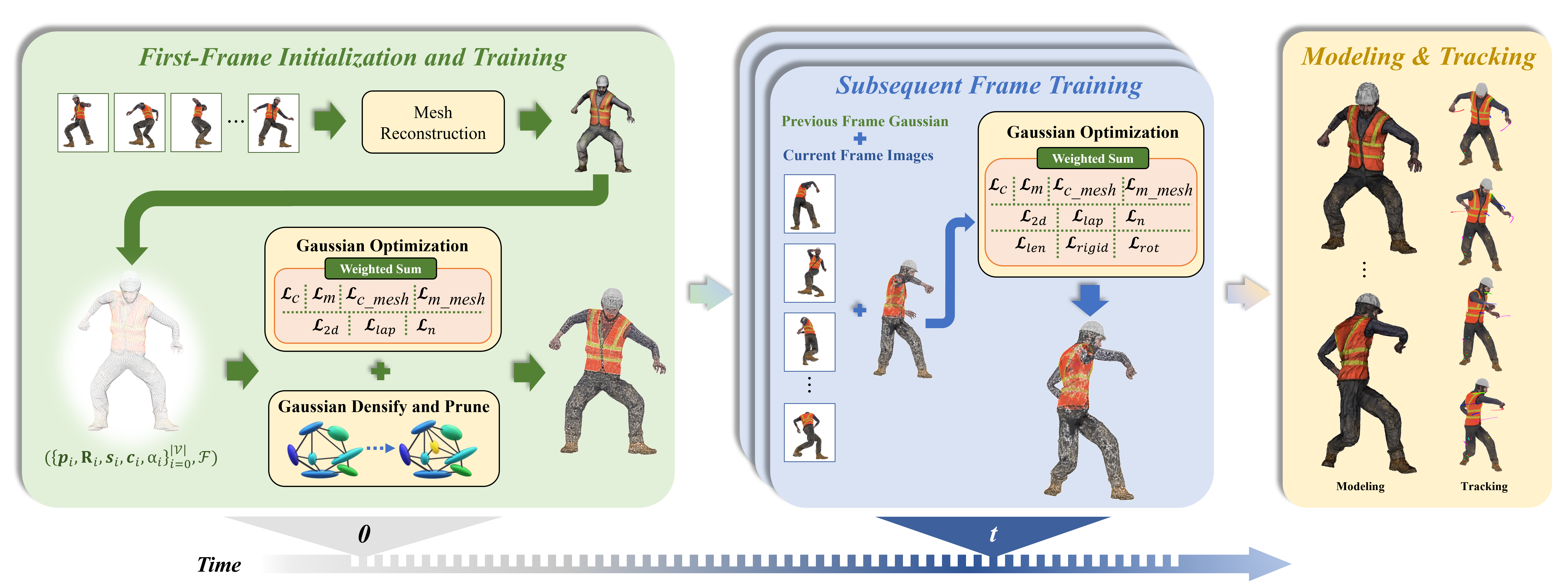}
    \vspace{-0.8cm}
    \caption{Overview of our pipeline. We first reconstruct an initial mesh from multi-view images and convert it into a Gaussian point cloud with manifold topology. A topology-aware densification and pruning strategy is then applied to refine the Gaussian representation while preserving surface connectivity. Temporal consistency constraints are introduced to enforce coherent deformation of Gaussians across frames. Finally, topology-consistent Gaussian and mesh sequences are obtained, enabling accurate 3D keypoint tracking.}
    \vspace{-0.6cm}
    \label{fig:pipeline}
\end{figure*}

Although Gaussian Splatting enables high-quality rendering, current animation production pipelines remain fundamentally mesh-based. Since Gaussian Splatting synthesizes images through 3D Gaussian functions, it still lacks mature solutions for lighting, deformation, and editing. Consequently, its integration into computer animation remains limited. Nonetheless, Gaussian Splatting provides a new perspective and valuable inspiration for advancing mesh reconstruction research.

\subsection{Mesh Reconstruction}

Mesh models have long been a central representation in 3D computer graphics. For dynamic scenes, reconstructing topology-consistent meshes is crucial for tasks such as motion control and model editing. Traditional NeRF-based methods\cite{park2021hypernerf, tang2023delicate, pumarola2021d, park2021nerfies} struggle to impose explicit topological constraints on dynamic objects, leading to reconstructions that lack temporal consistency. Although mesh models can be extracted for each frame using methods such as Marching Cubes \cite{lorense1987high} or Marching Tetrahedra\cite{treece1999regularised}, these approaches remain essentially frame-wise static reconstructions and do not ensure continuity across time. Recent works therefore aim to achieve high-fidelity modeling while preserving consistent spatial structure.

To address these issues, several Gaussian–based mesh reconstruction approaches have been proposed. Topo4D\cite{li2024topo4d} binds Gaussians to the vertices of a static template mesh and drives their positional deformation across frames while maintaining their topological relationships. However, this method is limited to head models and relies on a high-quality Metahuman template. GauSTAR\cite{zheng2025gaustar} reconstructs and tracks dynamic geometry using Gaussian Splatting guided by optical flow, yet its preprocessing pipeline is complex and difficult to deploy in production environments. More recent methods have made progress in both simplicity and reconstruction quality. DG-Mesh\cite{liu2024dynamic} attaches Gaussians to a mesh representation and can automatically reconstruct dynamic meshes, while Dynamic 2DGS\cite{zhang2024dynamic} incorporates 2D Gaussian primitives to achieve more accurate mesh reconstruction.

It is worth noting that although the aforementioned methods have made progress in temporal consistency and mesh reconstruction, they are still unable to produce topology-consistent mesh sequences. Most existing approaches reconstruct meshes by applying Poisson reconstruction or similar techniques to the point cloud of each frame. This makes it difficult to enforce consistent topology across frames, or even to maintain a fixed number of vertices. As a result, the reconstructed meshes are often unsuitable for downstream tasks such as animation production.

\section{Method}

To reconstruct a sequence of topology-consistent mesh models and enable robust keypoint tracking, we propose a topology-constrained dynamic Gaussian Splatting framework built upon standard 3D Gaussians (\cref{sec:3D Gaussian splatting}). Our pipeline starts by initializing Gaussians from the mesh reconstructed in the first frame. By introducing a manifold topology prior, we convert the initial mesh into a structured Gaussian point cloud 
. Based on this, we design a topology-aware Gaussian densification strategy that supports both training and refinement of Gaussians while preserving the underlying manifold structure(\cref{sec:Optimizing the Topological Structure of Gaussians}). For subsequent frames, we enforce temporal coherence of Gaussian parameters via 1-ring neighborhood regularization, ensuring consistent topology across the entire sequence (\cref{sec:Topology-Consistent Gaussian Training}). Once the dynamic Gaussians are fully optimized, we obtain a sequence of topology-consistent meshes and support stable 3D keypoint tracking (\cref{sec:Modeling and Tracking}). The overall pipeline is illustrated in the \cref{fig:pipeline}.

\subsection{3D Gaussian splatting}\label{sec:3D Gaussian splatting}

3D Gaussian splatting represent an explicit 3D representation method that is trained using multi-view images of the target object along with corresponding camera parameters. The target object is modeled by a large number of 3D Gaussians, each parameterized by its position $\boldsymbol{\mu}$, covariance matrix $\boldsymbol{\Sigma}$, and color $\boldsymbol{c}$. The Gaussian function is defined as follows:
\begin{equation}
  G(x) = {e^{ - \frac{1}{2}{{(\boldsymbol{x} - \boldsymbol{\mu} )}^T}\boldsymbol{\Sigma}^{-1} (\boldsymbol{x} - \boldsymbol{\mu} )}}
  \label{eq:Gx}
  \vspace{-0.1cm}
\end{equation}
 
The covariance matrix $\boldsymbol{\Sigma}$ must be positive semi-definite. Therefore, it is decomposed into two components: a rotation matrix $\boldsymbol{R}$ representing orientation, and a scaling matrix $\boldsymbol{S}$ representing scale. To ensure that the covariance matrix remains positive semi-definite during gradient descent optimization, it is defined as follows:
\begin{equation}
  \boldsymbol{\Sigma}  = \boldsymbol{RS}\boldsymbol{S}^T\boldsymbol{R}^T
  \label{eq:Sigma}
  \vspace{-0.2cm}
\end{equation}

For 2D image rendering, the color value $\boldsymbol{c_i}$ of each Gaussian at a pixel location is computed using spherical harmonics. The weight $\alpha_i$ is calculated based on opacity and the 2D Gaussian distribution. The pixel color $C$ is obtained by compositing $k$ 2D Gaussians in depth order as follows:

\begin{equation}
  C = \sum\limits_{i = 1}^{k} {{\alpha _i}} \prod\limits_{j = 1}^{i - 1} {(1 - {\alpha _j}){\boldsymbol{c}_i}} 
  \label{eq:C}
  \vspace{-0.2cm}
\end{equation}

\subsection{Topology-Aware Gaussian Initialization and Optimization}\label{sec:Optimizing the Topological Structure of Gaussians}

First, the topological structure of the Gaussian model needs to be optimized. To ensure correct topology reconstruction, this stage uses a canonical pose (e.g., A-pose or equivalent pose) as the initial reference. For the datasets we use, the first frame provides a suitable initial pose, enabling a stable topology optimization process.
We initialize Gaussian primitives from a high-quality multi-view reconstructed mesh of the first frame. Each Gaussian is associated with a vertex, inheriting its position, color, and connectivity to preserve topology. Rotation, scale, and opacity parameters are also initialized to roughly align the Gaussians with the mesh surface. This initialization provides a topology-consistent starting point for subsequent optimization, ensuring better convergence and reconstruction fidelity. 

\noindent\textbf{Gaussian Optimization}

To achieve accurate reconstruction in a canonical pose, we optimize several losses. Color losses $\mathcal{L}_{\text{c}}$ and $\mathcal{L}_{\text{mesh\_c}}$ align rendered images with ground truth. Mask losses $\mathcal{L}_{\text{m}}$ and $\mathcal{L}_{\text{mesh\_m}}$ enforce coverage of the target region. 2D scale loss $\mathcal{L}_{\text{2d}}$ improves Gaussian fitting to the object surface. Laplace smoothness $\mathcal{L}_{\text{lap}}$ reduces local spikes, and normal consistency $\mathcal{L}_{\text{n}}$ aligns Gaussian orientations with mesh normals. Joint optimization yields a detailed Gaussian representation with consistent topology.

Given the ground-truth image $\boldsymbol{I}_{gt}$ and the Gaussian Splatting rendering $\boldsymbol{I}_{gs}$, we use the 3DGS\cite{kerbl20233d} image loss.

\vspace{-0.5cm}
\begin{equation}
   {{\cal L}_{\rm{c}}^{gs}} = 0.8 \cdot \left\| {{{\boldsymbol{{I}}}_{gs}} - {{\boldsymbol{I}}_{gt}}} \right\| + 0.2 \cdot {{\cal L}_{ssim}}({{\boldsymbol{I}}_{gs}},{{\boldsymbol{I}}_{gt}})
  \label{eq:loss_color}
  \vspace{-0.15cm}
\end{equation}

We use a mask loss. ${{\boldsymbol{I}}_{m}}$ represents the rendering with Gaussian colors set to white and opacities set to 1, and ${{\boldsymbol{I}}{mask}}$ represents the ground-truth mask, which can be extracted using the SAM\cite{kirillov2023segment}. The loss is defined as:

\vspace{-0.5cm}
\begin{equation}
    {{\cal L}_{m}^{gs}} = 0.8 \cdot  \left\| {{{\boldsymbol{{I}}}_{m}} - {{\boldsymbol{I}}_{mask}}} \right\| + 0.2 \cdot {{\cal L}_{ssim}}({{\boldsymbol{I}}_{m}},{{\boldsymbol{I}}_{mask}})
  \label{eq:loss_mask}
  \vspace{-0.15cm}
\end{equation}

A mesh is constructed from the current Gaussian vertices and their topology, with Gaussian RGB values assigned as vertex colors. The mesh is rendered differentiably using Nvdiffrast\cite{laine2020modular}. The mesh losses $\mathcal{L}_{\text{c}}^{mesh}$ and $\mathcal{L}_{\text{m}}^{mesh}$ are computed analogously to the Gaussian image losses.

To enable Gaussian primitives to better represent the surface of the target object, we impose a constraint on the scale along the \(z\)-axis during training, encouraging it to be as small as possible. The smaller the scale along the \(z\)-axis, the closer the trained Gaussian model fits the target object shape\cite{li2024topo4d}. The scale constraint loss function we use is defined as:
\begin{equation}
  {{\cal L}_{2d}} = \sum\limits_{i=1}^k {s_z^i}
  \label{eq:loss_2dscale}
\end{equation}
where $k$ is the total number of Gaussians. ${s_z^i}$ represent the scale along the \(z\)-axis.

In addition to constraining the Gaussian \(z\)-axis scaling parameter, we adopt a Laplacian smoothness term as a regularization to penalize local spatial discontinuities\cite{liu2024dynamic}. This encourages smoother Gaussian positions and suppresses spiky artifacts by penalizing local spatial variations. 
\vspace{-0.1cm}
\begin{equation}
  {{\cal L}_{{\rm{lap}}}} = \frac{1}{k}\sum\limits_{i = 1}^k {{{\left\| {{\boldsymbol{\delta} _i}} \right\|}^2}} ,\quad{\boldsymbol{\delta} _i} = {\boldsymbol{\mu} _i} - \frac{1}{{\left| {{m}} \right|}}\sum\limits_{j=1}^{m} {{\boldsymbol{\mu} _j}}
  \label{eq:loss_smooth}
  \vspace{-0.1cm}
\end{equation}
where, $m$ denotes the number of 1-ring neighbors of Gaussian primitive $i$, and ${\mu}_i$ denotes its position.

We further introduce a normal consistency loss. For any Gaussian primitive, the vertex normal at the corresponding vertex, computed from the mesh model, is denoted as ${\boldsymbol{n}}^{mesh}$. We define the \(z\)-axis direction of the Gaussian’s rotation parameter as the Gaussian normal direction, denoted as ${\boldsymbol{n}}^{gs}$. The normal consistency loss is formulated as:

\vspace{-0.3cm}
\begin{equation}
  {{\cal L}_{n}} = \sum\limits_{i = 1}^k {{{\left\| {\boldsymbol{n}_i^{gs} - \boldsymbol{n}_i^{mesh}} \right\|}}}
  \label{eq:loss_normal}
\end{equation}
\vspace{-0.3cm}


Finally, all the aforementioned loss terms are combined through a weighted summation to construct the overall loss function for the first-frame Gaussian model, which guides the training of both Gaussian parameters and the topology structure.

\noindent\textbf{Topology-preserving densification and pruning}


Our method builds upon the original Gaussian densification and pruning strategies by introducing an update approach that maintains manifold topology. The original Gaussian method determines point addition or removal solely based on local properties of Gaussian points (e.g., projection gradients, opacity, scale). The newly added or removed points are treated independently without explicit connectivity between points. As a result, the manifold structure of the geometric surface is prone to disruption during iterations.

\begin{figure}[ht]
    \vspace{-0.3cm}
    \centering
    \includegraphics[width=\linewidth]{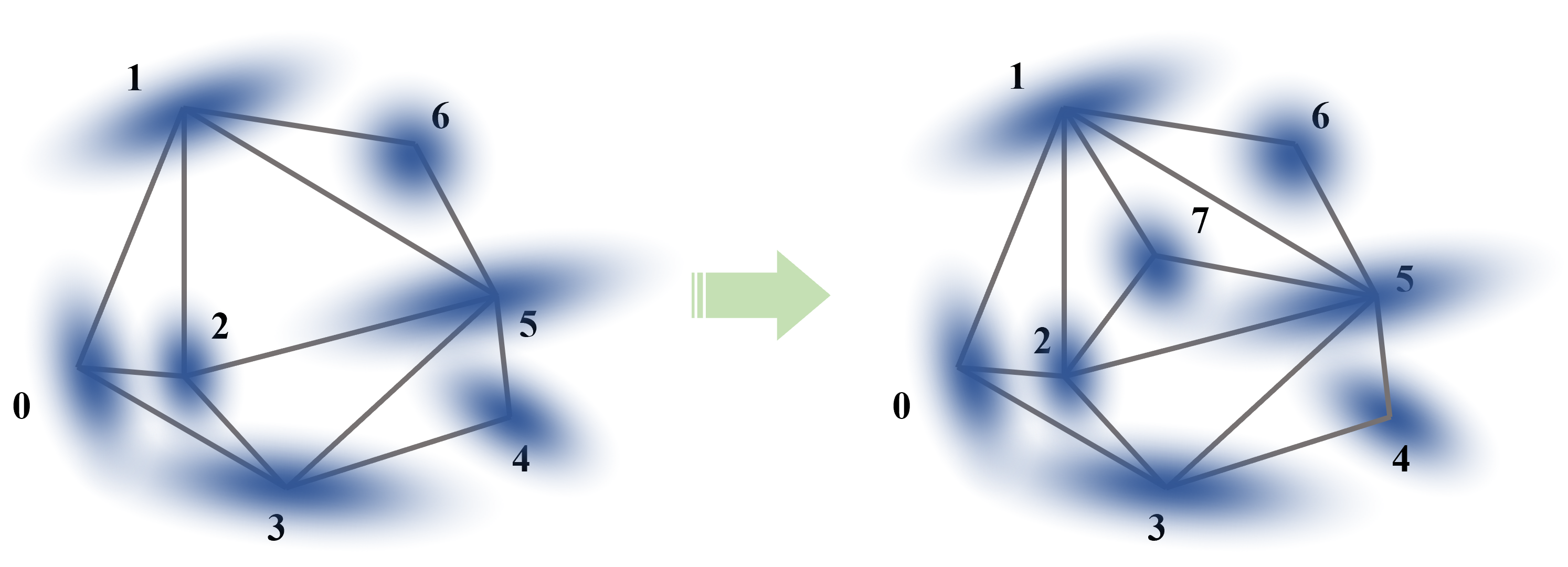}
    \vspace{-0.5cm}
    \caption{Topology-preserving densification.}
    \vspace{-0.2cm}
    \label{fig:densitify}

\end{figure}

To this end, while retaining the original Gaussian criteria for selecting points to densify or prune, we design a topology maintenance mechanism specifically for the addition and removal operations. The densification process is illustrated in \cref{fig:densitify}. During densification, for any triangle whose projection gradient exceeds a threshold, we insert a new Gaussian point in the parameter space. Its attributes are computed as the average of the three vertex Gaussian parameters. Simultaneously, the original triangle is subdivided into three new triangles: $(\mu_0, \mu_1, \mu_{\text{new}})$, $(\mu_1, \mu_2, \mu_{\text{new}})$, $(\mu_2, \mu_0, \mu_{\text{new}})$. The face set is updated accordingly, thereby maintaining the Gaussian topological connectivity during densification.
 
\begin{figure}
    \centering
    \vspace{-0.2cm}
    \includegraphics[width=\linewidth]{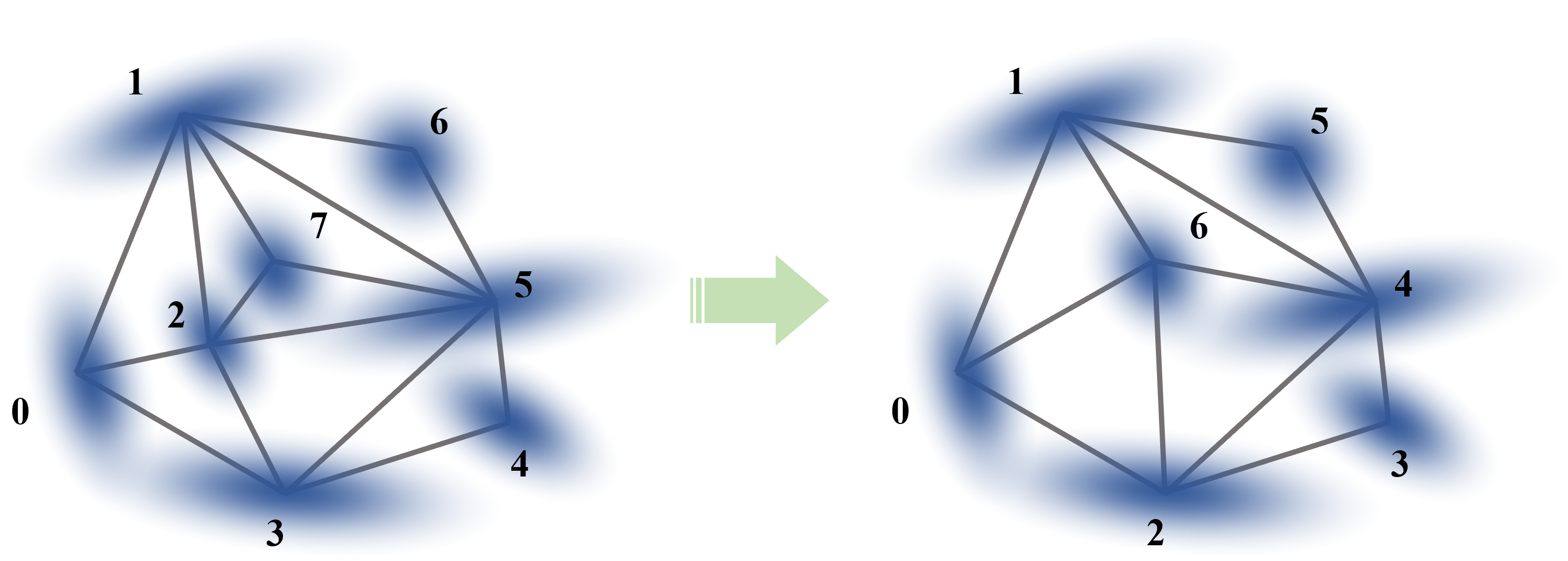}
    \vspace{-0.5cm}
    \caption{Topology-preserving pruning.}
    \vspace{-0.6cm}
    \label{fig:prune}
\end{figure}

During the pruning stage, we first identify Gaussian points to be removed based on the original criteria. Unlike the original gaussian splatting, we adopt a topology-preserving merging strategy with the nearest neighbor to maintain the manifold, as illustrated in \cref{fig:prune}. 
Inspired by the edge collapse strategy proposed by Hoppe et al.\cite{hoppe1999new}, we design an edge collapse cost function tailored for the Gaussian topology structure. The function integrates geometric and attribute errors in a weighted manner to minimize both geometric and visual loss during the simplification process. Specifically, for each Gaussian primitive to be pruned, we compute the collapse cost for all edges connecting it to its 1-ring neighboring primitives. The geometric error \(E_g\) is measured by the Euclidean distance between their centers, while the attribute error \(E_a\) is computed based on color differences. The overall collapse cost \(C\) is then obtained by combining the two:

\vspace{-0.3cm}
\begin{equation}
  C = \omega_g \cdot \| \mathbf{p}_i - \mathbf{p}_j \| + \omega_a \cdot \| \mathbf{c}_i - \mathbf{c}_j \|
  \label{eq:loss_normal}
\end{equation}
where \(\omega_g\) and \(\omega_a\) are the respective weighting coefficients used to unify the scales of different error terms. The edge with the minimal cost \(C\) is selected for collapse, thereby achieving efficient simplification while preserving topological consistency.


By incorporating this topology-aware update strategy into both densification and pruning, the Gaussian distribution is adaptively refined while the manifold structure of the surface is preserved throughout canonical-space optimization. This enables coherent and controllable geometric modeling. To maintain consistent topology during dynamic reconstruction, densification and pruning are applied only in the canonical space.

\begin{figure*}[h]
    \centering
    \vspace{-0.8cm}
    \includegraphics[width=\linewidth]{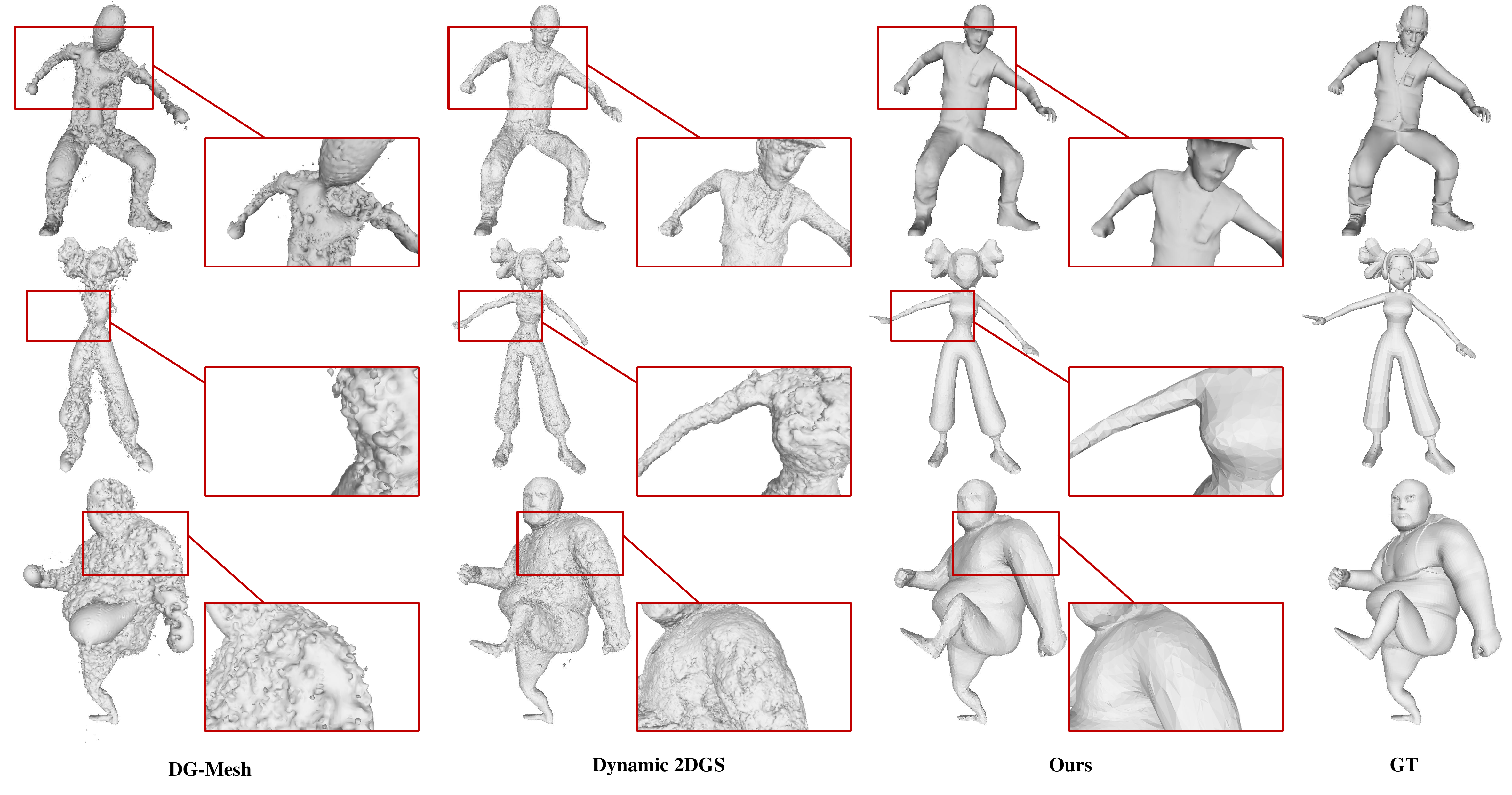}
    \vspace{-0.8cm}
    \caption{Mesh reconstruction comparison on the MIX-TAG dataset.}
    \vspace{-0.6cm}
    \label{fig:mix-tag mesh}
\end{figure*}

\subsection{Topology-Consistent Gaussian Training}\label{sec:Topology-Consistent Gaussian Training}

In the 3DGS training process, individual Gaussians have no explicit relationships. This leads to disordered outputs and uncontrolled relative positions, which compromise temporal consistency and continuity. To address this, we establish temporal consistency loss functions for subsequent frames. The edge length consistency loss $\mathcal{L}_{\text{len}}$ preserves the spatial structure. The rigidity constraint loss $\mathcal{L}_{\text{rigid}}$ leverages the 1-ring neighborhood to suppress local deformations. The rotation consistency loss $\mathcal{L}_{\text{rot}}$ ensures smooth rotational changes across frames. By jointly optimizing these losses, temporal coherence is maintained throughout multi-frame training.


To constrain the relative positions of Gaussians, we propose a Gaussian topology edge length consistency loss, which enforces continuity of distances between adjacent Gaussians across frames:
\vspace{-0.2cm}
\begin{equation}
  {{\cal L}_{len}} = \sum\limits_{i = 1}^{k_l} {\| {{l_{t,i}} - {l_{t - 1,i}}} \|}
  \label{eq:loss_edge}
\end{equation}

\vspace{-0.2cm}
\noindent where ${k_l}$ is the number of edges. For a Gaussian $i$, let ${l}_{t,i}$ denote the edge length at frame $t$ based on the topology. 

To better regulate the motion of Gaussian distributions and preserve topological information, we extend the Gaussian edge length loss. 
Following \cite{luiten2024dynamic}, we refine both the rigid loss and the rotation loss using the 1-ring neighborhood. The rigid loss is defined as follows:

\vspace{-0.6cm}

\begin{equation}
    {{\cal L}_{i,j}} = {{\| {( {\boldsymbol{\mu}_{j,t - 1} - \boldsymbol{\mu}_{i,t - 1}} ) - \Delta \boldsymbol{R}_t ( {\boldsymbol{\mu}_{j,t} - \boldsymbol{\mu}_{i,t}} )} \|}}
  \label{eq:loss_rigid2}
\end{equation}

\vspace{-0.45cm} 
\begin{equation}
  {{\cal L}_{rigid}} = \sum\limits_{i=1}^k {\sum\limits_{j=1}^{m} {{\omega _{i,j}} {\cal L}_{i,j} }}
  \label{eq:loss_rigid}
\end{equation}

\vspace{-0.1cm}
\noindent where $\Delta \boldsymbol{R}_t = {\boldsymbol{R}_{i,t - 1}}\boldsymbol{R}_{i,t}^{{\rm{ - 1}}}$, ${\omega _{i,j}} = \exp ( - {\lambda _w} \cdot {l_{i,j}})$, and $m$ is the number of 1-ring neighbors of \(\mu_i\). $\boldsymbol{R}_{i,t}$ denotes the rotation matrix of Gaussian primitive $i$ at frame $t$.



The rotation loss based on the 1-ring neighborhood is given by:
\vspace{-0.4cm}
\begin{equation}
  {{\cal L}_{rot}} = \sum\limits_{i=1}^k {\sum\limits_{j=1}^{m} {{\omega _{i,j}}\left\| {{\boldsymbol{q}_{j,t}} \otimes \boldsymbol{q}_{j,t-1}^{{\rm{ - 1}}} - {\boldsymbol{q}_{i,t}} \otimes \boldsymbol{q}_{i,t-1}^{{\rm{ - 1}}}} \right\|} }
  \label{eq:loss_rot}
  \vspace{-0.3cm}
\end{equation}

\noindent where $\boldsymbol{q}$ denotes the normalized quaternion. $\boldsymbol{q}^{\rm{ - 1}}$ denotes the conjugate of the quaternion. $\otimes$ denotes quaternion multiplication.

The loss function for training Gaussians in subsequent frames is a weighted sum of the three topology constraints above and the first frame training loss. 

\subsection{Modeling and Tracking}\label{sec:Modeling and Tracking}


After training, we obtain a sequence of Gaussians with consistent topology. The position parameters of each Gaussian serve as the vertex positions of the mesh, while the learned manifold topology defines the connectivity. Using the Gaussian positions as vertices, a mesh can be generated for each frame. By first reducing the 3D spherical harmonic parameters to a single dimension and then converting it to RGB values, colored mesh models can be exported from Gaussian Splatting.

Any 3D point on the reconstructed mesh surface can be represented by a set of barycentric coordinates and a triangle face index. Since we obtain a sequence of meshes with consistent topology, the parameterization of these points applies to the mesh models of all frames. Thus, the trajectory of any target point can be tracked across all frames using this representation.

\section{Experiments}

\begin{table*}[h]
  \centering
  \vspace{-0.7cm}
  \begin{tabular}{lcccccccc}
    \toprule
     &  &PSNR$_{gs}\uparrow$    &SSIM$_{gs}\uparrow$    &LPIPS$_{gs}\downarrow$ &CD$\downarrow$    &EMD$\downarrow$   & Tracking MSE$\downarrow$ & T-C Mesh \\
    \midrule

            & Dynamic 3DGS    & 30.56& 0.97& 0.026&\redcross& \redcross& 0.000676& \redcross\\
            & DG-Mesh     & 22.46& 0.95& 0.080& 1.48& 0.29& 0.013502& \redcross\\
    Boxer   & Deformable-GS & 34.38& \textbf{0.98}& 0.016& \redcross& \redcross& 0.006631& \redcross\\   
            & Dynamic 2DGS  & 34.27& 0.97& 0.018& 0.47& 0.13& 0.009148& \redcross\\    
            & Ours & \textbf{34.76}& \textbf{0.98}& \textbf{0.15}& \textbf{0.32}& \textbf{0.010}& \textbf{0.000569} & \greencheck\\    
    \midrule   
            & Dynamic 3DGS    & 30.19& 0.97& 0.021& \redcross& \redcross& 0.000329& \redcross\\
            & DG-Mesh     & 19.27& 0.94& 0.093& 10.31& 0.43  & 0.019222& \redcross\\
    Dancer   & Deformable-GS & 19.08& 0.94& 0.081& \redcross& \redcross& 0.042462& \redcross\\   
            & Dynamic 2DGS  & 27.45& \textbf{0.98}& 0.030& 0.75& 0.13  & 0.042436& \redcross\\    
            & Ours & \textbf{34.61}& \textbf{0.98}& \textbf{0.010}& \textbf{0.24}& \textbf{0.088} & \textbf{0.000101}& \greencheck\\    
    \midrule   
            & Dynamic 3DGS    & 29.22& 0.95& 0.031 & \redcross& \redcross& 0.000226 & \redcross\\
            & DG-Mesh     & 19.13& 0.90& 0.11 & 7.15& 0.35 & 0.048107 & \redcross\\
    Worker   & Deformable-GS & 31.93& 0.96& 0.026 & \redcross& \redcross& 0.028535 & \redcross\\   
            & Dynamic 2DGS  & 31.86& 0.96& 0.023 & 0.45& 0.12 & 0.057903 & \redcross\\    
            & Ours & \textbf{32.17}& \textbf{0.97}& \textbf{0.021} & \textbf{0.39}& \textbf{0.10} & \textbf{0.000218} & \greencheck\\        
    \bottomrule
  \end{tabular}
  \vspace{-0.2cm}
  \caption{Rendering, modeling, and tracking quality comparison on the MIX-TAG dataset. T-C Mesh indicates whether a method can reconstruct a topology-consistent mesh sequence.}
  \vspace{-0.7cm}
  \label{tab:MIX-TAG render}
\end{table*}

\subsection{Data Preparation}
We evaluated our algorithm on both synthetic and real datasets. First, we rendered a dataset named MIX-TAG using motion data downloaded from Mixamo to evaluate modeling and tracking performance. The images in the dataset have a resolution of 1080×1080, with each subject captured from 42 different viewpoints. MIX-TAG provides ground-truth mesh, enabling quantitative evaluation of keypoint tracking accuracy and mesh reconstruction accuracy. For real data, we trained and tested on the TalkBody4D dataset, introduced in Taoavatar \cite{chen2025taoavatar}. It contains 4D human motion data captured by 59 precisely calibrated RGB cameras at 20 FPS, with a resolution of 3000×4000 pixels. In our experiments, we used only 30 of these camera views for training and evaluation. Our algorithm is implemented in PyTorch and trained on an NVIDIA RTX 3090 GPU with 24GB of memory. For the initial mesh reconstruction, we adopt the NeuS2\cite{wang2023neus2} method.

\subsection{Comparison}

\begin{figure}[h]
    \centering
    \includegraphics[width=\linewidth]{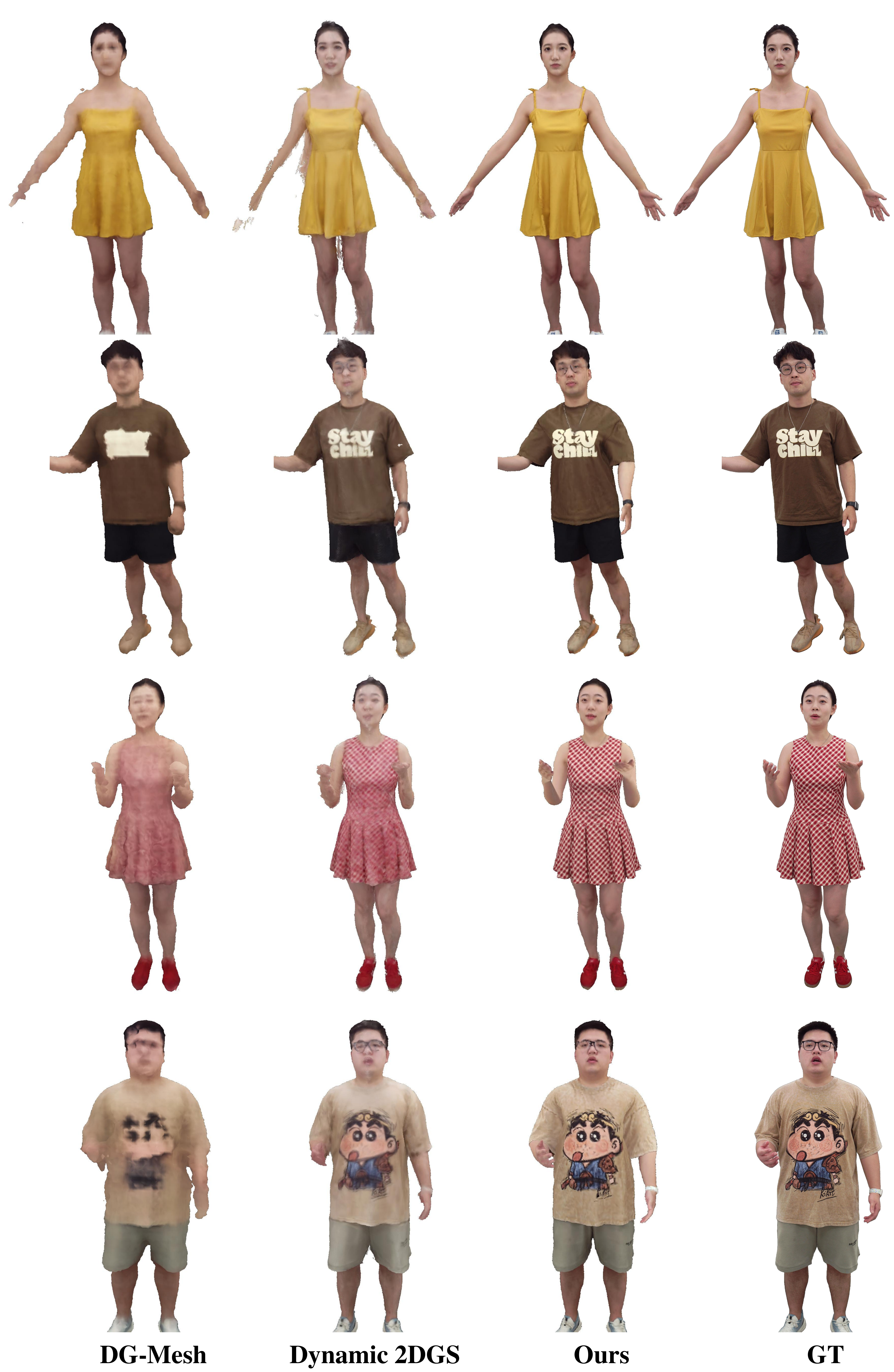}
    \vspace{-0.5cm}
    \caption{Mesh rendering results on the TalkBody4D dataset.}
    \vspace{-0.6cm}
    \label{fig:TalkBody4D render}
\end{figure}

\begin{figure}[h]
    \centering
    \includegraphics[width=\linewidth]{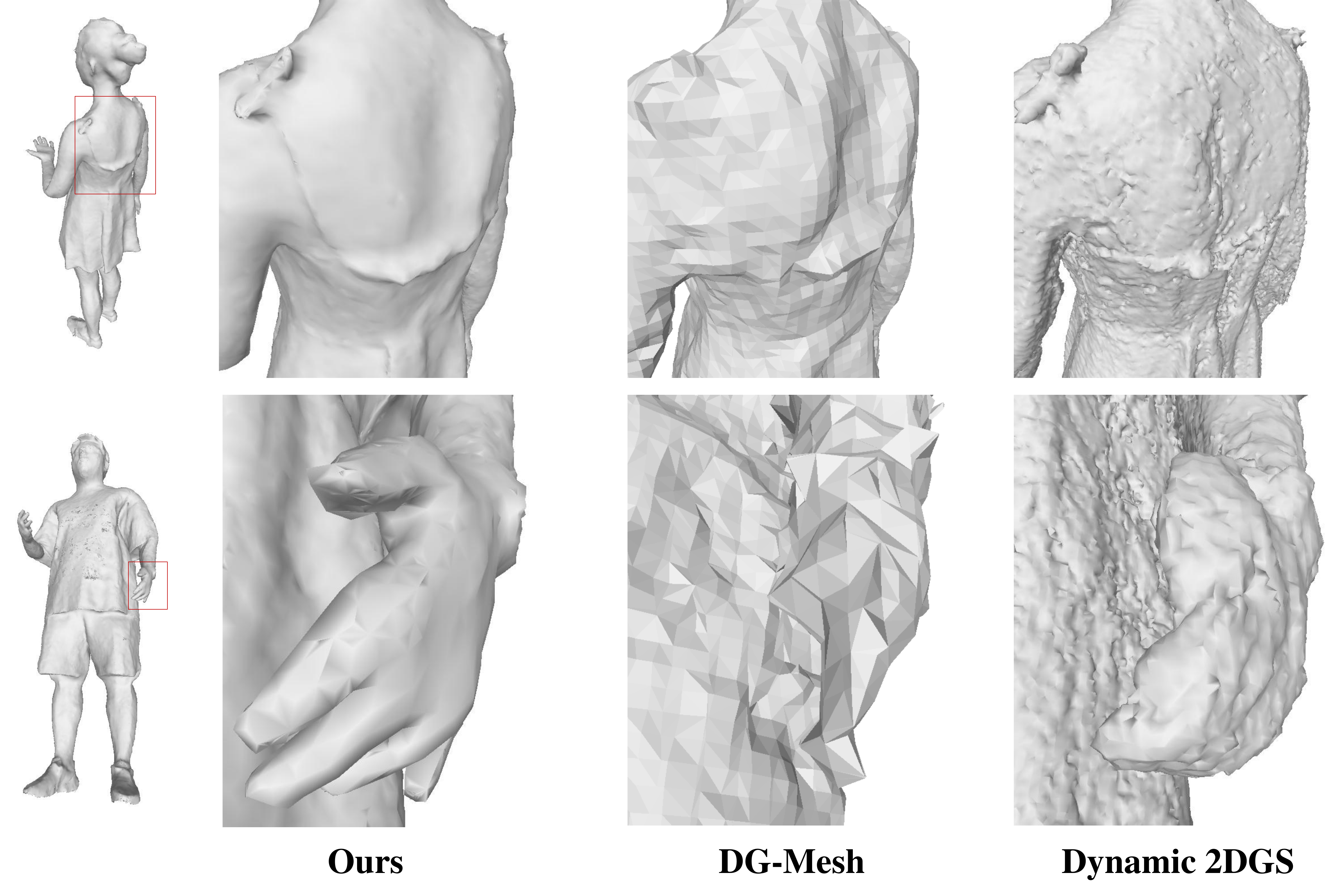}
    \vspace{-0.8cm}
    \caption{Our method produces smoother and more realistic surface geometry while maintaining topological consistency.}
    \vspace{-0.4cm}
    \label{fig:TalkBody4D mesh}
\end{figure}

\begin{figure}[h]
    \centering
    \includegraphics[width=\linewidth]{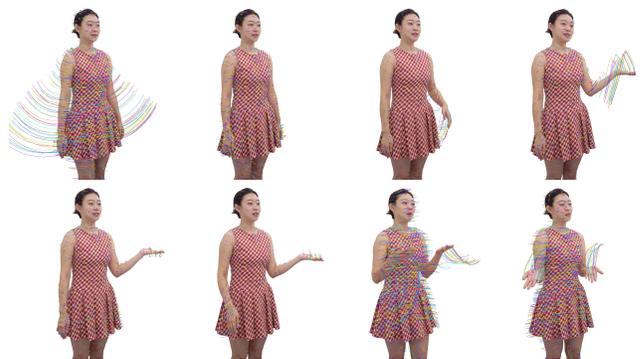}
    \vspace{-0.7cm}
    \caption{Tracking results on the TalkBody4D dataset. Our method produces stable and accurate keypoint trajectories.}
    \vspace{-0.5cm}
    \label{fig:TalkBody4D tracking}
\end{figure}

\begin{figure*}[ht]
    \centering
    \vspace{-0.7cm}
    \includegraphics[width=\linewidth]{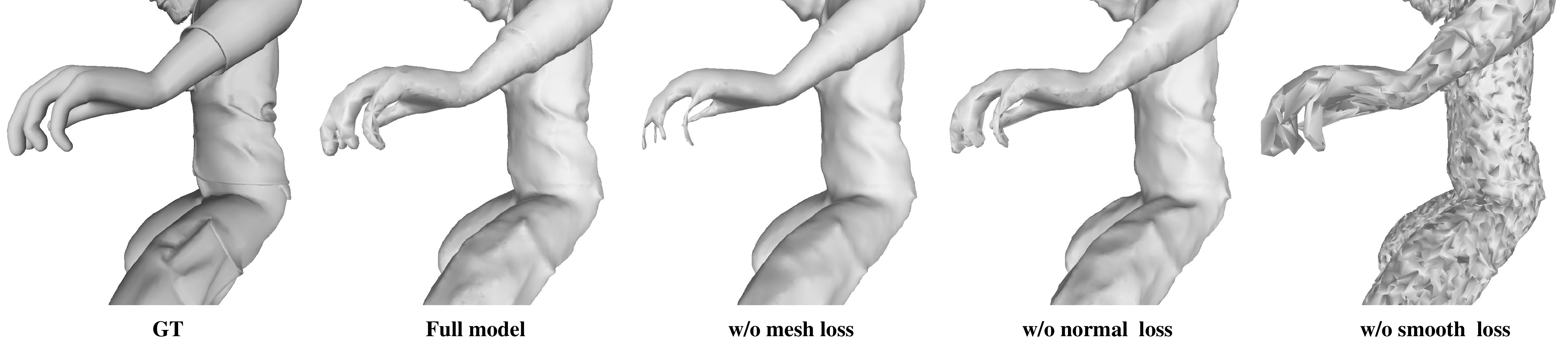}
    \vspace{-0.7cm}
    \caption{Ablation study on mesh reconstruction. Visual comparison shows that mesh loss, normal loss, and smoothness loss all contribute to more accurate and artifact-free reconstructions.}
    \vspace{-0.6cm}
    \label{fig:ablation mesh}
\end{figure*}




We conducted comparative experiments on reconstruction accuracy and tracking precision using the MIX-TAG dataset. This dataset includes arm motions, leg motions, and full-body motions of human subjects with diverse skin tones and body types. 
The \cref{tab:MIX-TAG render} presents a quality comparison of the Gaussian rendering results.

For mesh quality evaluation, we use Chamfer Distance and Earth Mover’s Distance to measure the reconstruction error relative to the ground-truth mesh. As shown in the \cref{fig:mix-tag mesh}, our method produces meshes with smoother and more realistic vertex distributions compared to other approaches. Compared with other methods, our approach is distinguished by producing reconstructed mesh sequences that preserve the consistent topology. As shown in \cref{tab:MIX-TAG render}, where our method outperforms the baselines.

For tracking accuracy, we processed the original mesh models from the MIX-TAG dataset and recorded motion sequences of representative mesh vertices as ground-truth trajectories. We use Mean Squared Error (MSE) to measure the difference between the algorithm’s tracking results and the ground truth, representing tracking accuracy. The tracking accuracy results are reported in the \cref{tab:MIX-TAG render}.


We also evaluated reconstruction results on TalkBody4D\cite{chen2025taoavatar} dataset. Since this dataset does not provide ground-truth mesh models, we cannot quantitatively compare reconstruction or tracking accuracy. The mesh rendering results are shown in the \cref{fig:TalkBody4D render}, with detailed metrics reported in the \cref{tab:TalkBody4D render}.

We compared the reconstructed meshes; the reconstruction results and details are presented in the \cref{fig:TalkBody4D mesh}, showing that our method generates more realistic and smoother meshes while maintaining the consistent topology.
We visualized tracking results on TalkBody4D, as shown in the \cref{fig:TalkBody4D tracking}. It can be observed that our method produces  stable tracking. 

\begin{table}[htb]
  \centering
  \vspace{-0.3cm}
  \begin{tabular}{lcccc}
    \toprule
    Method & PSNR$_m\uparrow$ & SSIM$_m\uparrow$  & LPIPS$_m\downarrow$ \\
    \midrule

    DG-Mesh         &24.90 &0.91 &0.097	\\
    Dynamic-2DGS    &23.58 &0.90 &0.081	\\
    Ours            &\textbf{26.82} &\textbf{0.93} &\textbf{0.077}	\\

    \bottomrule
  \end{tabular}
  \vspace{-0.2cm}
  \caption{Mesh rendering quality on the TalkBody4D dataset.}
  \vspace{-0.4cm}
  \label{tab:TalkBody4D render}
\end{table}

\subsection{Ablations}
We conducted ablation studies on the MIX-TAG dataset to analyze the contribution of different losses and structural components. 
We first validated the effectiveness of various static reconstruction losses and the proposed topology-preserving densification and pruning (D\&P) module, with quantitative results summarized in \cref{tab:ablation mesh0}. As shown in \cref{fig:ablation mesh}, removing the smooth loss leads to surface artifacts, while the mesh and normal losses are crucial for accurate geometry recovery. In addition, removing the D\&P module results in degraded reconstruction fidelity, verifying its role in maintaining manifold topology and preserving fine surface details.
We further evaluated the temporal consistency losses, with results reported in \cref{tab:ablation mesh1}. The results show that the length, rigidity, and rotation constraints jointly ensure smooth deformation across frames. Removing any of these terms causes noticeable degradation in both reconstruction accuracy and tracking stability, confirming their complementary effects.


\begin{table}[h]
  \centering
  \footnotesize
  \begin{tabular}{lcccc}
    \toprule
    Method  &PSNR$_{gs}\uparrow$  &PSNR$_m\uparrow$    & CD$\downarrow$    & EMD$\downarrow$ \\
    \midrule
    Full model	        &\textbf{32.90}	&\textbf{29.75}	&\textbf{0.360}	&\textbf{0.10703}\\
    w/o mesh loss       &27.70	&27.41	&0.396	&0.11238\\
    w/o normal loss     &27.42	&28.21	&0.384	&0.11038\\
    w/o smooth loss	    &28.82	&26.55	&0.408	&0.11231\\
    w/o D\&P$^{\dagger}$ &30.22	&28.68	&0.378	&0.10984\\
    \bottomrule
  \end{tabular}
  \vspace{-0.2cm}
  \caption{Ablation study of static reconstruction losses. Removing mesh, normal, or smoothness loss degrades reconstruction fidelity and mesh quality. $^{\dagger}$ D\&P denotes the topology-preserving densification and pruning strategy.\textbf{}}
  \vspace{-0.2cm}
  \label{tab:ablation mesh0}
\end{table}

\begin{table}[h]
  \centering
  \begin{tabular}{lc}
    \toprule
    Method  &Tracking MSE$\downarrow$ \\
    \midrule

    Full model	     &\textbf{0.000368}\\
    w/o length loss     &0.000412\\
    w/o rigid loss	 &0.000395\\
    w/o rot loss     &0.000372\\

    \bottomrule
  \end{tabular}
  \vspace{-0.2cm}
  \caption{Ablation study of temporal consistency loss functions. Removing rigid, rotation, or isotropic constraints reduces tracking accuracy.}
  \vspace{-0.6cm}
  \label{tab:ablation mesh1}
\end{table}


\section{Conclusion}

We present a topology-aware dynamic reconstruction and tracking framework based on 3D Gaussian Splatting. The method reconstructs topology-consistent meshes and supports precise 3D keypoint tracking across dynamic sequences. Our method bridges the gap between Gaussian-based and mesh-based representations for dynamic reconstruction. The topology-preserving densification and pruning method ensures the manifold topology while increasing the number of Gaussians. Furthermore, a series of regularization terms is introduced to achieve temporally smooth deformation of Gaussian parameters while preserving topology. Experimental results demonstrate that our method outperforms existing approaches on both synthetic and real datasets, achieving significant improvements in mesh reconstruction accuracy and keypoint tracking precision. This provides a high-fidelity and cost-effective solution for downstream tasks such as animation production.

\textbf{Limitation:} Our method demonstrates strong performance in dynamic mesh modeling and tracking, effectively handling non-rigid deformations while maintaining consistent topology. However, it should be noted that the current framework is primarily designed for scenarios with stable topological relationships. It lacks sufficient adaptability to handle cases with drastic topological changes, such as clothing tears or object splits. This limitation restricts its applicability in more complex dynamic scenes. Future research could explore extending the method to accommodate dynamic objects with changing topology. We believe progress in this direction will broaden the method’s applicability and increase its practical value.

{
    \small
    \bibliographystyle{ieeenat_fullname}
    \bibliography{main}

@String(ECCV= {Eur. Conf. Comput. Vis.})

@String(TOG= {ACM Trans. Graph.})

@String(ECCV  = {ECCV})

@String(TOG   = {ACM TOG})

@article{mildenhall2021nerf,
  title={Nerf: Representing scenes as neural radiance fields for view synthesis},
  author={Mildenhall, Ben and Srinivasan, Pratul P and Tancik, Matthew and Barron, Jonathan T and Ramamoorthi, Ravi and Ng, Ren},
  journal={Communications of the ACM},
  volume={65},
  number={1},
  pages={99--106},
  year={2021},
  publisher={ACM New York, NY, USA}
}

@article{muller2022instant,
  title={Instant neural graphics primitives with a multiresolution hash encoding},
  author={M{\"u}ller, Thomas and Evans, Alex and Schied, Christoph and Keller, Alexander},
  journal={ACM transactions on graphics (TOG)},
  volume={41},
  number={4},
  pages={1--15},
  year={2022},
  publisher={ACM New York, NY, USA}
}

@inproceedings{fridovich2022plenoxels,
  title={Plenoxels: Radiance fields without neural networks},
  author={Fridovich-Keil, Sara and Yu, Alex and Tancik, Matthew and Chen, Qinhong and Recht, Benjamin and Kanazawa, Angjoo},
  booktitle={Proceedings of the IEEE/CVF conference on computer vision and pattern recognition},
  pages={5501--5510},
  year={2022}
}

@article{wang2021neus,
  title={Neus: Learning neural implicit surfaces by volume rendering for multi-view reconstruction},
  author={Wang, Peng and Liu, Lingjie and Liu, Yuan and Theobalt, Christian and Komura, Taku and Wang, Wenping},
  journal={arXiv preprint arXiv:2106.10689},
  year={2021}
}

@article{yariv2021volume,
  title={Volume rendering of neural implicit surfaces},
  author={Yariv, Lior and Gu, Jiatao and Kasten, Yoni and Lipman, Yaron},
  journal={Advances in neural information processing systems},
  volume={34},
  pages={4805--4815},
  year={2021}
}

@article{kerbl20233d,
  title={3D Gaussian splatting for real-time radiance field rendering.},
  author={Kerbl, Bernhard and Kopanas, Georgios and Leimk{\"u}hler, Thomas and Drettakis, George},
  journal={ACM Trans. Graph.},
  volume={42},
  number={4},
  pages={139--1},
  year={2023}
}

@inproceedings{luiten2024dynamic,
  title={Dynamic 3d gaussians: Tracking by persistent dynamic view synthesis},
  author={Luiten, Jonathon and Kopanas, Georgios and Leibe, Bastian and Ramanan, Deva},
  booktitle={2024 International Conference on 3D Vision (3DV)},
  pages={800--809},
  year={2024},
  organization={IEEE}
}

@article{park2021hypernerf,
  title={HyperNeRF: a higher-dimensional representation for topologically varying neural radiance fields},
  author={Park, Keunhong and Sinha, Utkarsh and Hedman, Peter and Barron, Jonathan T and Bouaziz, Sofien and Goldman, Dan B and Martin-Brualla, Ricardo and Seitz, Steven M},
  journal={ACM Transactions on Graphics (TOG)},
  volume={40},
  number={6},
  pages={1--12},
  year={2021},
  publisher={ACM New York, NY, USA}
}

@inproceedings{li2024topo4d,
  title={Topo4D: Topology-Preserving Gaussian Splatting for High-fidelity 4D Head Capture},
  author={Li, Xuanchen and Cheng, Yuhao and Ren, Xingyu and Jia, Haozhe and Xu, Di and Zhu, Wenhan and Yan, Yichao},
  booktitle={European Conference on Computer Vision},
  pages={128--145},
  year={2024},
  organization={Springer}
}

@inproceedings{zheng2025gaustar,
  title={GauSTAR: Gaussian Surface Tracking and Reconstruction},
  author={Zheng, Chengwei and Xue, Lixin and Zarate, Juan and Song, Jie},
  booktitle={Proceedings of the Computer Vision and Pattern Recognition Conference},
  pages={16543--16553},
  year={2025}
}

@inproceedings{liu2024dynamic,
  title={Dynamic Gaussians Mesh: Consistent Mesh Reconstruction from Dynamic Scenes},
  author={Liu, Isabella and Su, Hao and Wang, Xiaolong},
  year={2024},
  booktitle={The Thirteenth International Conference on Learning Representations}
}

@inproceedings{wang2023neus2,
  title={Neus2: Fast learning of neural implicit surfaces for multi-view reconstruction},
  author={Wang, Yiming and Han, Qin and Habermann, Marc and Daniilidis, Kostas and Theobalt, Christian and Liu, Lingjie},
  booktitle={Proceedings of the IEEE/CVF International Conference on Computer Vision},
  pages={3295--3306},
  year={2023}
}

@article{dongye2024lodavatar,
  title={Lodavatar: Hierarchical embedding and adaptive levels of detail with gaussian splatting for enhanced human avatars},
  author={Dongye, Xiaonuo and Guo, Hanzhi and Luo, Le and Jiang, Haiyan and Bao, Yihua and Tian, Zeyu and Weng, Dongdong},
  journal={arXiv preprint arXiv:2410.20789},
  year={2024}
}

@article{chen2025ps,
  title={PS-GS: Gaussian Splatting for Multi-View Photometric Stereo},
  author={Chen, Yixiao and Liang, Bin and Guo, Hanzhi and Cheng, Yongqing and Zhao, Jiayi and Weng, Dongdong},
  journal={arXiv preprint arXiv:2507.18231},
  year={2025}
}

@inproceedings{gaussian_grouping,
    title={Gaussian Grouping: Segment and Edit Anything in 3D Scenes},
    author={Ye, Mingqiao and Danelljan, Martin and Yu, Fisher and Ke, Lei},
    booktitle={ECCV},
    year={2024}
}

@inproceedings{relightable,
  title={Relightable gaussian codec avatars},
  author={Saito, Shunsuke and Schwartz, Gabriel and Simon, Tomas and Li, Junxuan and Nam, Giljoo},
  booktitle={Proceedings of the IEEE/CVF Conference on Computer Vision and Pattern Recognition},
  pages={130--141},
  year={2024}
}

@article{jung2023deformable,
  title={Deformable 3D Gaussian Splatting for Animatable Human Avatars},
  author={Jung, HyunJun and Brasch, Nikolas and Song, Jifei and P{\'e}rez-Pellitero, Eduardo and Zhou, Yiren and Li, Zhihao and Navab, Nassir and Busam, Benjamin},
  journal={CoRR},
  year={2023}
}

@inproceedings{gaussianheadavatar,
            title={Gaussian Head Avatar: Ultra High-fidelity Head Avatar via Dynamic Gaussians},
            author={Xu, Yuelang and Chen, Benwang and Li, Zhe and Zhang, Hongwen and Wang, Lizhen and Zheng, Zerong and Liu, Yebin},
            booktitle={Proceedings of the IEEE/CVF Conference on Computer Vision and Pattern Recognition},
            year={2024}
          }

@inproceedings{jiang2024hifi4g,
  title={Hifi4g: High-fidelity human performance rendering via compact gaussian splatting},
  author={Jiang, Yuheng and Shen, Zhehao and Wang, Penghao and Su, Zhuo and Hong, Yu and Zhang, Yingliang and Yu, Jingyi and Xu, Lan},
  booktitle={Proceedings of the IEEE/CVF conference on computer vision and pattern recognition},
  pages={19734--19745},
  year={2024}
}

@article{zhang2024dynamic,
  title={Dynamic 2d gaussians: Geometrically accurate radiance fields for dynamic objects},
  author={Zhang, Shuai and Wu, Guanjun and Xie, Zhoufeng and Wang, Xinggang and Feng, Bin and Liu, Wenyu},
  journal={arXiv preprint arXiv:2409.14072},
  year={2024}
}

@article{jiang2025topology,
  title={Topology-Aware Optimization of Gaussian Primitives for Human-Centric Volumetric Videos},
  author={Jiang, Yuheng and Guo, Chengcheng and Wu, Yize and Hong, Yu and Zhu, Shengkun and Shen, Zhehao and Zhang, Yingliang and Jiao, Shaohui and Su, Zhuo and Xu, Lan and others},
  journal={arXiv preprint arXiv:2509.07653},
  year={2025}
}

@inproceedings{chen2025taoavatar,
  title={TaoAvatar: Real-Time Lifelike Full-Body Talking Avatars for Augmented Reality via 3D Gaussian Splatting},
  author={Chen, Jianchuan and Hu, Jingchuan and Wang, Gaige and Jiang, Zhonghua and Zhou, Tiansong and Chen, Zhiwen and Lv, Chengfei},
  booktitle={Proceedings of the Computer Vision and Pattern Recognition Conference},
  pages={10723--10734},
  year={2025}
}

@inproceedings{huang20242d,
  title={2d gaussian splatting for geometrically accurate radiance fields},
  author={Huang, Binbin and Yu, Zehao and Chen, Anpei and Geiger, Andreas and Gao, Shenghua},
  booktitle={ACM SIGGRAPH 2024 conference papers},
  pages={1--11},
  year={2024}
}

@inproceedings{xue2023nsf,
  title={Nsf: Neural surface fields for human modeling from monocular depth},
  author={Xue, Yuxuan and Bhatnagar, Bharat Lal and Marin, Riccardo and Sarafianos, Nikolaos and Xu, Yuanlu and Pons-Moll, Gerard and Tung, Tony},
  booktitle={Proceedings of the IEEE/CVF international conference on computer vision},
  pages={15049--15060},
  year={2023}
}

@article{hanocka2020point2mesh,
  title={Point2Mesh},
  author={Hanocka, Rana and Metzer, Gal and Giryes, Raja and Cohen-Or, Daniel},
  journal={ACM Transactions on Graphics},
  volume={39},
  number={4},
  year={2020},
  publisher={Association for Computing Machinery (ACM)}
}

@article{collet2015high,
  title={High-quality streamable free-viewpoint video},
  author={Collet, Alvaro and Chuang, Ming and Sweeney, Pat and Gillett, Don and Evseev, Dennis and Calabrese, David and Hoppe, Hugues and Kirk, Adam and Sullivan, Steve},
  journal={ACM Transactions on Graphics (ToG)},
  volume={34},
  number={4},
  pages={1--13},
  year={2015},
  publisher={ACM New York, NY, USA}
}

@article{chen2023neusg,
  title={Neusg: Neural implicit surface reconstruction with 3d gaussian splatting guidance},
  author={Chen, Hanlin and Li, Chen and Lee, Gim Hee},
  journal={arXiv preprint arXiv:2312.00846},
  year={2023}
}

@inproceedings{guedon2024sugar,
  title={Sugar: Surface-aligned gaussian splatting for efficient 3d mesh reconstruction and high-quality mesh rendering},
  author={Gu{\'e}don, Antoine and Lepetit, Vincent},
  booktitle={Proceedings of the IEEE/CVF Conference on Computer Vision and Pattern Recognition},
  pages={5354--5363},
  year={2024}
}

@inproceedings{tang2023delicate,
  title={Delicate textured mesh recovery from nerf via adaptive surface refinement},
  author={Tang, Jiaxiang and Zhou, Hang and Chen, Xiaokang and Hu, Tianshu and Ding, Errui and Wang, Jingdong and Zeng, Gang},
  booktitle={Proceedings of the IEEE/CVF International Conference on Computer Vision},
  pages={17739--17749},
  year={2023}
}

@inproceedings{mescheder2019occupancy,
  title={Occupancy networks: Learning 3d reconstruction in function space},
  author={Mescheder, Lars and Oechsle, Michael and Niemeyer, Michael and Nowozin, Sebastian and Geiger, Andreas},
  booktitle={Proceedings of the IEEE/CVF conference on computer vision and pattern recognition},
  pages={4460--4470},
  year={2019}
}

@inproceedings{park2019deepsdf,
  title={Deepsdf: Learning continuous signed distance functions for shape representation},
  author={Park, Jeong Joon and Florence, Peter and Straub, Julian and Newcombe, Richard and Lovegrove, Steven},
  booktitle={Proceedings of the IEEE/CVF conference on computer vision and pattern recognition},
  pages={165--174},
  year={2019}
}

@inproceedings{hoppe1999new,
  title={New quadric metric for simplifying meshes with appearance attributes},
  author={Hoppe, Hugues},
  booktitle={Proceedings Visualization'99 (Cat. No. 99CB37067)},
  pages={59--510},
  year={1999},
  organization={IEEE}
}

@inproceedings{xie2024physgaussian,
  title={Physgaussian: Physics-integrated 3d gaussians for generative dynamics},
  author={Xie, Tianyi and Zong, Zeshun and Qiu, Yuxing and Li, Xuan and Feng, Yutao and Yang, Yin and Jiang, Chenfanfu},
  booktitle={Proceedings of the IEEE/CVF Conference on Computer Vision and Pattern Recognition},
  pages={4389--4398},
  year={2024}
}

@article{abou2024physically,
  title={Physically embodied gaussian splatting: A realtime correctable world model for robotics},
  author={Abou-Chakra, Jad and Rana, Krishan and Dayoub, Feras and S{\"u}nderhauf, Niko},
  journal={arXiv preprint arXiv:2406.10788},
  year={2024}
}

@article{choi2024phys3dgs,
  title={Phys3DGS: Physically-based 3D Gaussian Splatting for Inverse Rendering},
  author={Choi, Euntae and Yoo, Sungjoo},
  journal={arXiv preprint arXiv:2409.10335},
  year={2024}
}

@inproceedings{hu2024gaussianavatar,
  title={Gaussianavatar: Towards realistic human avatar modeling from a single video via animatable 3d gaussians},
  author={Hu, Liangxiao and Zhang, Hongwen and Zhang, Yuxiang and Zhou, Boyao and Liu, Boning and Zhang, Shengping and Nie, Liqiang},
  booktitle={Proceedings of the IEEE/CVF conference on computer vision and pattern recognition},
  pages={634--644},
  year={2024}
}

@inproceedings{li2024animatable,
  title={Animatable gaussians: Learning pose-dependent gaussian maps for high-fidelity human avatar modeling},
  author={Li, Zhe and Zheng, Zerong and Wang, Lizhen and Liu, Yebin},
  booktitle={Proceedings of the IEEE/CVF conference on computer vision and pattern recognition},
  pages={19711--19722},
  year={2024}
}

@inproceedings{qian2024gaussianavatars,
  title={Gaussianavatars: Photorealistic head avatars with rigged 3d gaussians},
  author={Qian, Shenhan and Kirschstein, Tobias and Schoneveld, Liam and Davoli, Davide and Giebenhain, Simon and Nie{\ss}ner, Matthias},
  booktitle={Proceedings of the IEEE/CVF Conference on Computer Vision and Pattern Recognition},
  pages={20299--20309},
  year={2024}
}

@inproceedings{moreau2024human,
  title={Human gaussian splatting: Real-time rendering of animatable avatars},
  author={Moreau, Arthur and Song, Jifei and Dhamo, Helisa and Shaw, Richard and Zhou, Yiren and P{\'e}rez-Pellitero, Eduardo},
  booktitle={Proceedings of the IEEE/CVF conference on computer vision and pattern recognition},
  pages={788--798},
  year={2024}
}

@article{liu2025compgs++,
  title={CompGS++: Compressed Gaussian Splatting for Static and Dynamic Scene Representation},
  author={Liu, Xiangrui and Wu, Xinju and Wang, Shiqi and Li, Zhu and Kwong, Sam},
  journal={arXiv preprint arXiv:2504.13022},
  year={2025}
}

@inproceedings{liu2024compgs,
  title={Compgs: Efficient 3d scene representation via compressed gaussian splatting},
  author={Liu, Xiangrui and Wu, Xinju and Zhang, Pingping and Wang, Shiqi and Li, Zhu and Kwong, Sam},
  booktitle={Proceedings of the 32nd ACM International Conference on Multimedia},
  pages={2936--2944},
  year={2024}
}

@article{chen2025pcgs,
  title={Pcgs: Progressive compression of 3d gaussian splatting},
  author={Chen, Yihang and Li, Mengyao and Wu, Qianyi and Lin, Weiyao and Harandi, Mehrtash and Cai, Jianfei},
  journal={arXiv preprint arXiv:2503.08511},
  year={2025}
}

@inproceedings{chen2024hac,
  title={Hac: Hash-grid assisted context for 3d gaussian splatting compression},
  author={Chen, Yihang and Wu, Qianyi and Lin, Weiyao and Harandi, Mehrtash and Cai, Jianfei},
  booktitle={European Conference on Computer Vision},
  pages={422--438},
  year={2024},
  organization={Springer}
}

@article{chen2025hac++,
  title={Hac++: Towards 100x compression of 3d gaussian splatting},
  author={Chen, Yihang and Wu, Qianyi and Lin, Weiyao and Harandi, Mehrtash and Cai, Jianfei},
  journal={arXiv preprint arXiv:2501.12255},
  year={2025}
}

@article{laine2020modular,
  title={Modular primitives for high-performance differentiable rendering},
  author={Laine, Samuli and Hellsten, Janne and Karras, Tero and Seol, Yeongho and Lehtinen, Jaakko and Aila, Timo},
  journal={ACM Transactions on Graphics (ToG)},
  volume={39},
  number={6},
  pages={1--14},
  year={2020},
  publisher={ACM New York, NY, USA}
}

@article{lorense1987high,
  title={A high resolution 3D surface construction algorithm},
  author={Lorense, WE},
  journal={Proc. Siggraph 87},
  year={1987}
}

@article{treece1999regularised,
  title={Regularised marching tetrahedra: improved iso-surface extraction},
  author={Treece, Graham M and Prager, Richard W and Gee, Andrew H},
  journal={Computers \& Graphics},
  volume={23},
  number={4},
  pages={583--598},
  year={1999},
  publisher={Elsevier}
}

@inproceedings{kirillov2023segment,
  title={Segment anything},
  author={Kirillov, Alexander and Mintun, Eric and Ravi, Nikhila and Mao, Hanzi and Rolland, Chloe and Gustafson, Laura and Xiao, Tete and Whitehead, Spencer and Berg, Alexander C and Lo, Wan-Yen and others},
  booktitle={Proceedings of the IEEE/CVF international conference on computer vision},
  pages={4015--4026},
  year={2023}
}

@inproceedings{pumarola2021d,
  title={D-nerf: Neural radiance fields for dynamic scenes},
  author={Pumarola, Albert and Corona, Enric and Pons-Moll, Gerard and Moreno-Noguer, Francesc},
  booktitle={Proceedings of the IEEE/CVF conference on computer vision and pattern recognition},
  pages={10318--10327},
  year={2021}
}

@inproceedings{park2021nerfies,
  title={Nerfies: Deformable neural radiance fields},
  author={Park, Keunhong and Sinha, Utkarsh and Barron, Jonathan T and Bouaziz, Sofien and Goldman, Dan B and Seitz, Steven M and Martin-Brualla, Ricardo},
  booktitle={Proceedings of the IEEE/CVF international conference on computer vision},
  pages={5865--5874},
  year={2021}
}
}

\clearpage
\setcounter{page}{1}
\maketitlesupplementary

\renewcommand{\thesection}{\Alph{section}}
\setcounter{section}{0}
\setcounter{equation}{0}
\setcounter{figure}{0}
\setcounter{table}{0}

\section{Initialization of Gaussian Primitives} \label{sec:sup_A}

The initialization of Gaussian primitives is a crucial step that determines the effectiveness of subsequent optimization. Since Gaussian rendering optimization is inherently non-convex, the quality of the initial Gaussian model largely affects the reconstruction fidelity and convergence speed. In recent years, multi-view mesh reconstruction has developed several mature methods. To obtain a high-quality initial topology, we first perform multi-view geometric reconstruction on the first frame. This produces a dense 3D mesh model. The Gaussian model is then initialized based on the reconstructed mesh.

Let the reconstructed mesh be denoted as $\mathcal{M} = (\mathcal{V}, \mathcal{F})$, where $\mathcal{V} = \{\boldsymbol{v}_i \in \mathbb{R}^3\}$ is the set of vertices and $\mathcal{F} = \{\boldsymbol{f}_i \in \mathbb{Z}^3\}$ is the set of faces. We initialize the position parameter $\boldsymbol{p}_i \in \mathbb{R}^3$ of each Gaussian primitive by the corresponding vertex position $\boldsymbol{v}_i$, that is, $\boldsymbol{p}_i = \boldsymbol{v}_i$.

The color parameter $\boldsymbol{c}_i \in \mathbb{R}^3$ is directly initialized using the RGB color of the corresponding mesh vertex. The opacity parameter $\alpha_i$ is uniformly initialized to a constant value of 1. The rotation parameter $\boldsymbol{R}_i \in SO(3)$ is initialized based on the vertex normal $\boldsymbol{n}_i$. Specifically, $\boldsymbol{n}_i$ is used as the local coordinate system’s $z$-axis direction.

For the scale parameter $\boldsymbol{s}_i = (s_{i,x}, s_{i,y}, s_{i,z})^\top$, we initialize the $s_{i,x}$ and $s_{i,y}$ based on the distance to neighboring vertices to ensure coverage. Meanwhile, the the scale factor $s_{i,z}$ along the surface normal direction is set to a small value $\epsilon$, so that the Gaussian primitive closely adheres to the surface:
\begin{equation}
  s_{i,z} = \epsilon \ll s_{i,x}, s_{i,y}
  \label{eq:scale}
\end{equation}

Finally, the initial set of Gaussians can be represented as $\mathcal{G} = \left( \left\{ \left(\boldsymbol{p}_i, \boldsymbol{R}_i, \boldsymbol{s}_i, \boldsymbol{c}_i, \alpha_i\right) \right\}_{i=1}^{|\mathcal{V}|}, \mathcal{F} \right)$. Here, $\mathcal{F}$ represents the face index information inherited from the mesh model, which defines the topological connections between vertices. This topology can be employed in designing consistency regularization, facilitating the preservation of local geometric coherence of the target surface during Gaussian optimization.

\section{Detail of topology-preserving densification and pruning process}\label{sec:sup_B}

We propose a method that preserves the manifold topology of the model during the densification and pruning processes. The goal is to balance geometric detail and structural quality. Specifically, the topology-preserving densification strategy (see \cref{alg:densify}) enhances geometric details in sparse regions to improve reconstruction realism, while the topology-preserving pruning strategy (see \cref{alg:prune}) removes redundant Gaussian primitives and maintains the integrity of the final topology. Working together, these two processes ensure the generation of high-quality mesh models with consistent topology.

\begin{algorithm}[t]
\caption{Topology-preserving Gaussian Densification}
\label{alg:densify}
\KwIn{Gaussian scales $S$, other Gaussian parameters $P$, topology $\mathcal{T}$, vertex gradients $G$, gradients thresholds $\tau_g$, scale thresholds $\tau_s$}
\KwOut{Updated $\mathcal{T}$, $S$, $P$}

\ForEach{face $f=(v_1,v_2,v_3)\in\mathcal{T}$}{
    $g_f \leftarrow \frac{1}{3}(G_{v_1}+G_{v_2}+G_{v_3})$
    
    \If{$g_f > \tau_g$}{
        $\bar{s} \leftarrow \frac{1}{3}(S_{v_1}+S_{v_2}+S_{v_3})$
        
        \uIf{$\bar{s}<\tau_s$}{
            $s_{new} \leftarrow \bar{s}$
        }\Else{
            $s_{new} \leftarrow \frac{1}{2}\bar{s}$
            
            $S_{v_i} \leftarrow \frac{1}{2}S_{v_i},~i=1,2,3$
        }
        $p_{new} \leftarrow \frac{1}{3}(P_{v_1}+P_{v_2}+P_{v_3})$
        
        $\mathcal{T}\leftarrow \mathcal{T}\cup\text{new\_connections}$
        
        $S\leftarrow S\cup s_{new},~P\leftarrow P\cup p_{new}$
    }
}
\end{algorithm}







\begin{algorithm}[t]
\caption{Topology-preserving Gaussian Pruning (simplified)}
\label{alg:prune}
\KwIn{Gaussian parameters $\mathcal{G}$, topology $\mathcal{T}$, opacity threshold $\epsilon$}
\KwOut{Updated $\mathcal{G}$ and $\mathcal{T}$}

\ForEach{$g_i \in \mathcal{G}$}{
    \If{$\alpha_i < \epsilon$ \textbf{or} \text{IsTooLarge}($g_i$)}{
        $\mathcal{N}_i \leftarrow$ 1-ring neighbors of $g_i$ in $\mathcal{T}$

        $p \leftarrow \arg\min_{g_j \in \mathcal{N}_i} \text{CollapseCost}(g_i, g_j)$ 

        $\mathcal{T} \leftarrow$ EdgeCollapse($\mathcal{T}$, $p$)

        $\mathcal{G} \gets$ Remove($\mathcal{G}$, $g_i$)
    }
}

\end{algorithm}

\section{Loss function weights} \label{sec:sup_C}

We designed two sets of loss functions for the first frame and the subsequent frames. Each set combines multiple sub-losses with weighted sums to balance different optimization objectives. The specific weight configurations used during training are summarized in the \cref{tab:weights}.

\begin{table}[h]
  \centering
  \begin{tabular}{lc}
    \toprule
    Loss & Weight \\
    \midrule

    $\mathcal{L}_{\text{c}}^{gs}$            &1.0 	\\
    $\mathcal{L}_{\text{c}}^{mesh}$      &1.0  \\
    $\mathcal{L}_{\text{m}}^{gs}$            &3.0 \\
    $\mathcal{L}_{\text{m}}^{mesh}$      &3.0 \\
    $\mathcal{L}_{\text{2d}}$           &1.0 \\
    $\mathcal{L}_{\text{lap}}$          &5.0 \\
    $\mathcal{L}_{\text{n}}$            &1.0 \\
    $\mathcal{L}_{\text{len}}$          &4.0 \\
    $\mathcal{L}_{\text{rigid}}$        &4.0 \\
    $\mathcal{L}_{\text{rot}}$          &20.0 \\
    
    \bottomrule
  \end{tabular}
  \caption{Loss Function Weights.}
  \label{tab:weights}
\end{table}

\section{Introduction to the MIX-TAG Dataset} \label{sec:sup_C}

The MIX-TAG dataset comprises three dynamic human subjects: Worker, Dancer, and Boxer. To construct the dataset, we developed an automated data acquisition pipeline in Blender, deploying 42 uniformly distributed virtual cameras in each scene to perform frame-by-frame multi-view rendering of the animation sequences. In addition to RGB images, we simultaneously generated accurate binary masks for each frame. All rendered images have a resolution of 1080×1080.

In terms of sequence composition, Worker contains a 90-frame animation, while Dancer and Boxer each consist of 130 frames. These sequences include a wide range of challenging motion patterns, such as full-body coordinated motion, large leg swings, and torso rotations, ensuring the diversity and comprehensiveness of the dataset.


\begin{figure}[h]
    \centering
    \vspace{0.8cm}
    \includegraphics[width=\linewidth]{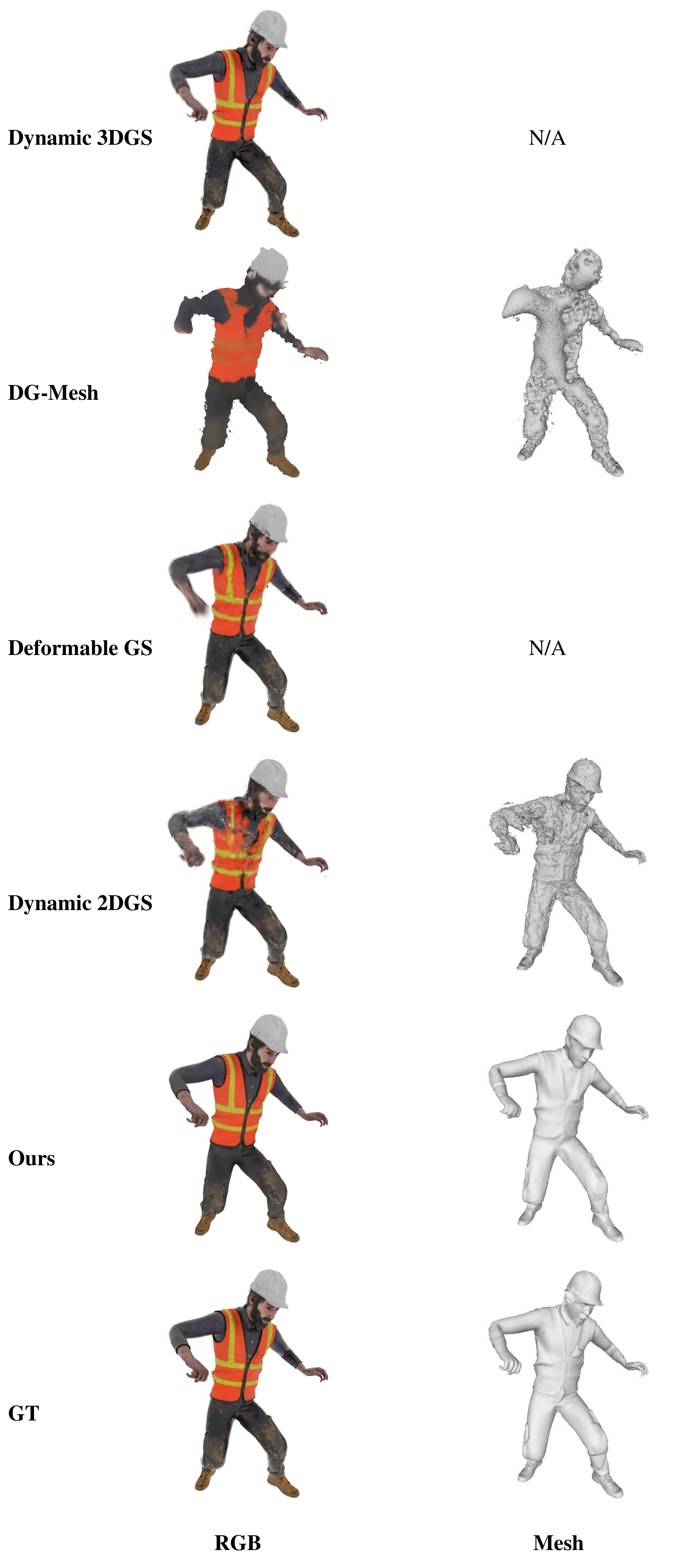}
    \caption{Comparison of training results for the Worker object.}
    \vspace{0.5cm}
    \label{fig:mix-tag_sup_0}
\end{figure}

\begin{figure}[h]
    \centering
    \vspace{0.8cm}
    \includegraphics[width=\linewidth]{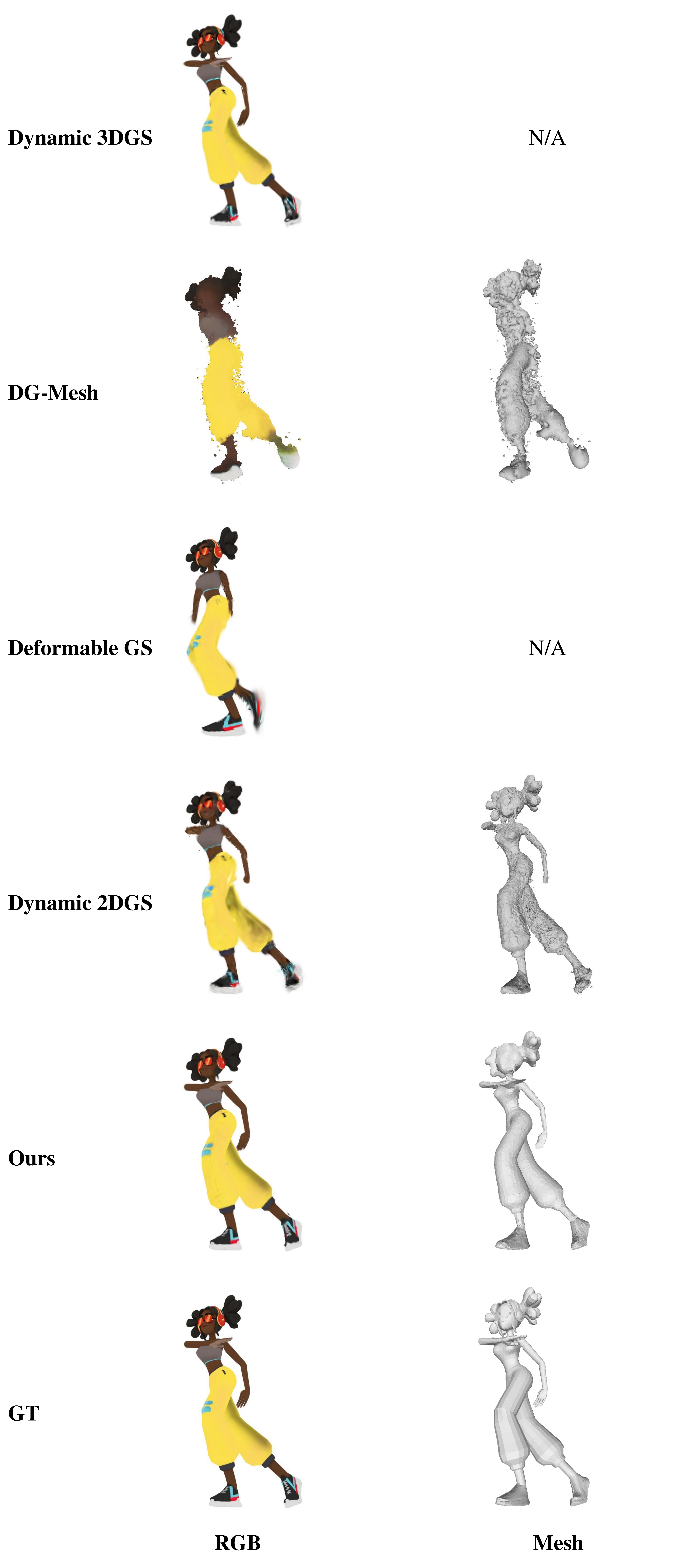}
    \caption{Comparison of training results for the Dancer object.}
    \vspace{0.5cm}
    \label{fig:mix-tag_sup_1}
\end{figure}

\begin{figure}[h]
    \centering
    \vspace{0.8cm}
    \includegraphics[width=\linewidth]{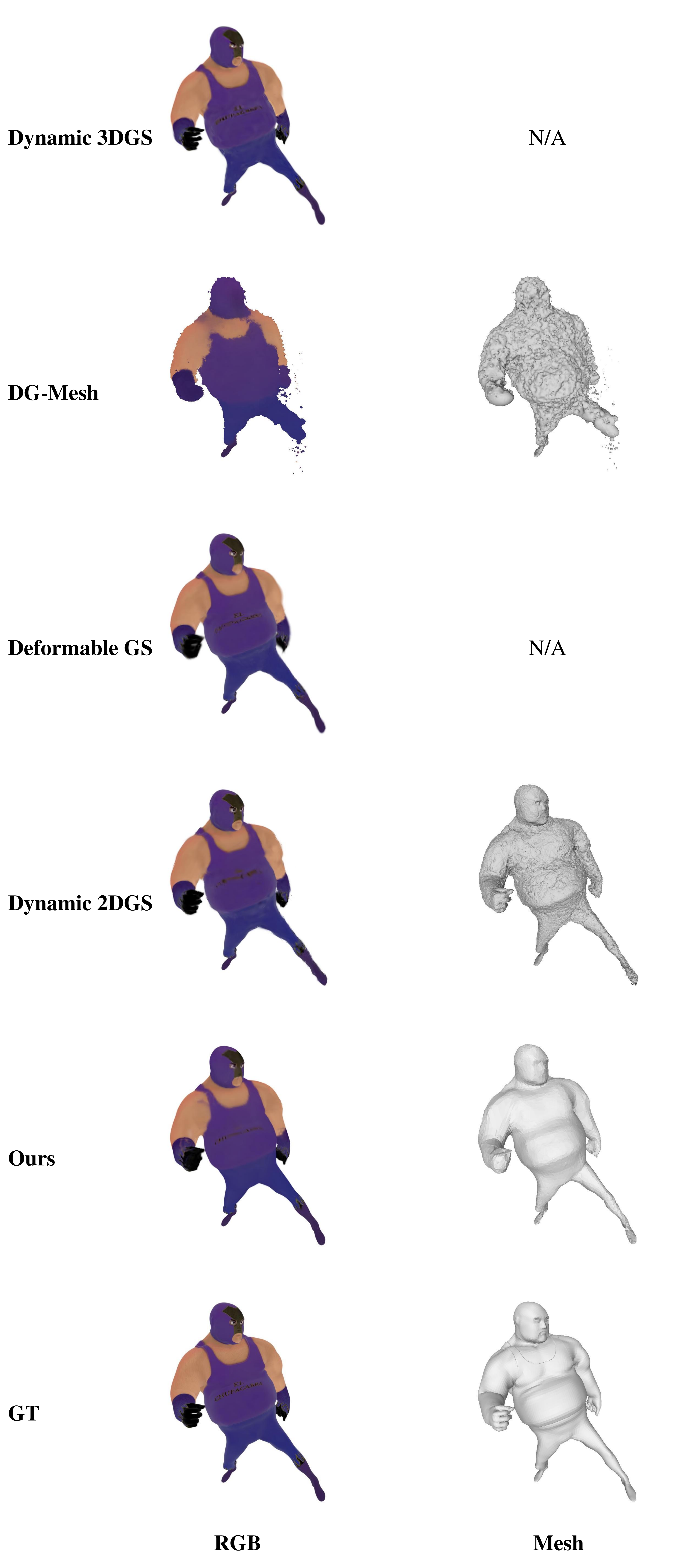}
    \caption{Comparison of training results for the Boxer object.}
    \vspace{0.5cm}
    \label{fig:mix-tag_sup_2}
\end{figure}

\begin{figure*}
    \centering
    \vspace{0.8cm}
    \includegraphics[width=\linewidth]{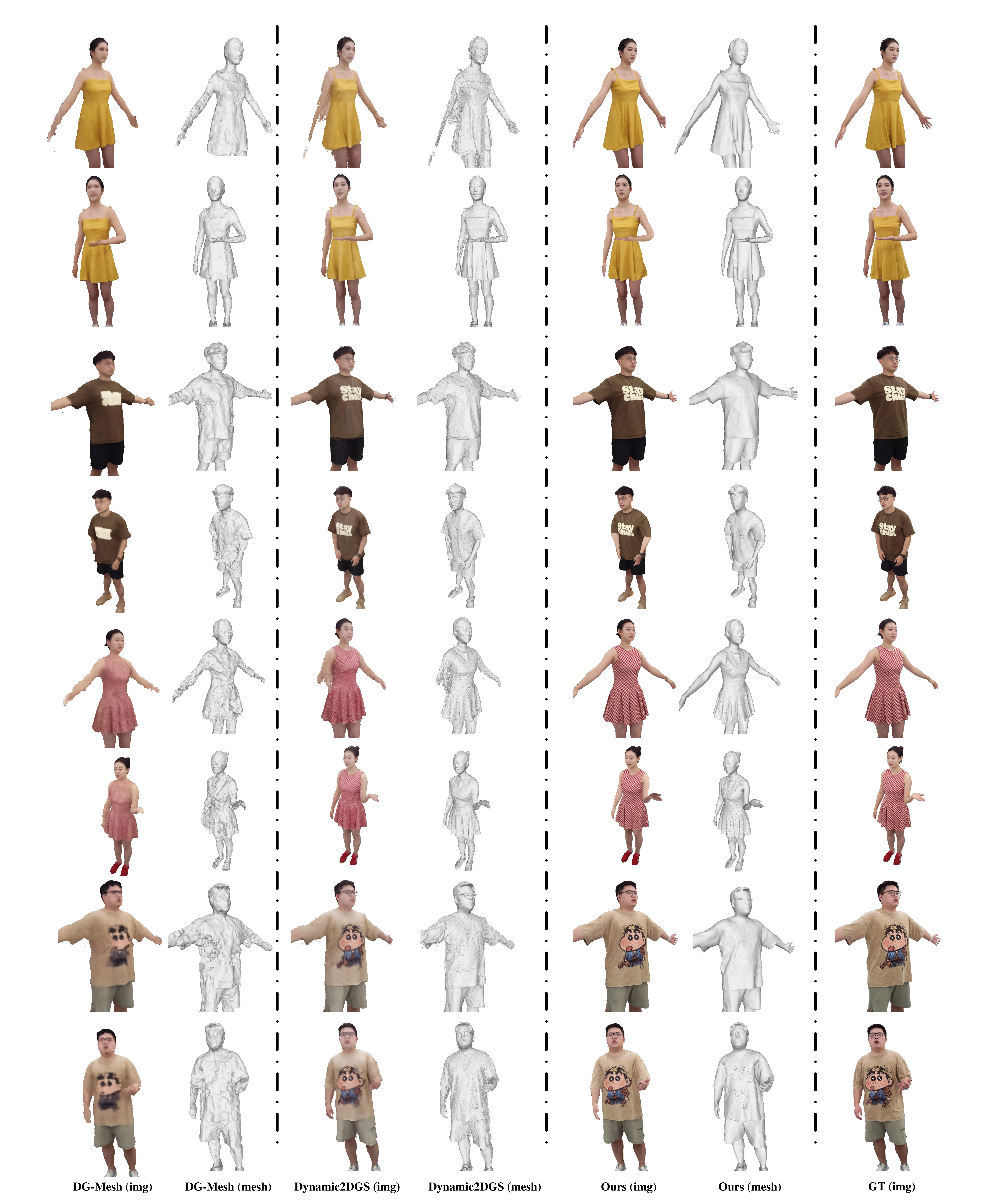}
    \caption{Comparison of training results on Talkbody4D dataset.}
    \vspace{0.5cm}
    \label{fig:ali_sup}
\end{figure*}

\section{More results on the MIX-TAG dataset} \label{sec:sup_E}

\cref{fig:mix-tag_sup_0}, \cref{fig:mix-tag_sup_1} and \cref{fig:mix-tag_sup_2} present both the Gaussian rendering results and the mesh reconstruction results obtained from models trained on the MIX-TAG dataset. Our method is compared with several Gaussian-based reconstruction approaches, including Dynamic 3DGS, DG-Mesh, Deformable-GS, and Dynamic 2DGS. The experimental results clearly demonstrate that, compared with existing baselines, our method achieves superior performance in both mesh reconstruction accuracy and rendering quality.

\section{More results on the Talkbody4D dataset} \label{sec:sup_E}

\cref{fig:ali_sup} presents additional results on the real dataset. We compare our method with DG-Mesh and Dynamic 2DGS in terms of mesh rendering and mesh reconstruction. The results show that our approach produces meshes with higher geometric fidelity, better structural coherence, and smoother surfaces. Furthermore, the meshes render results exhibit more realistic appearance compared with existing methods.

\begin{figure}
    \centering
    \includegraphics[width=\linewidth]{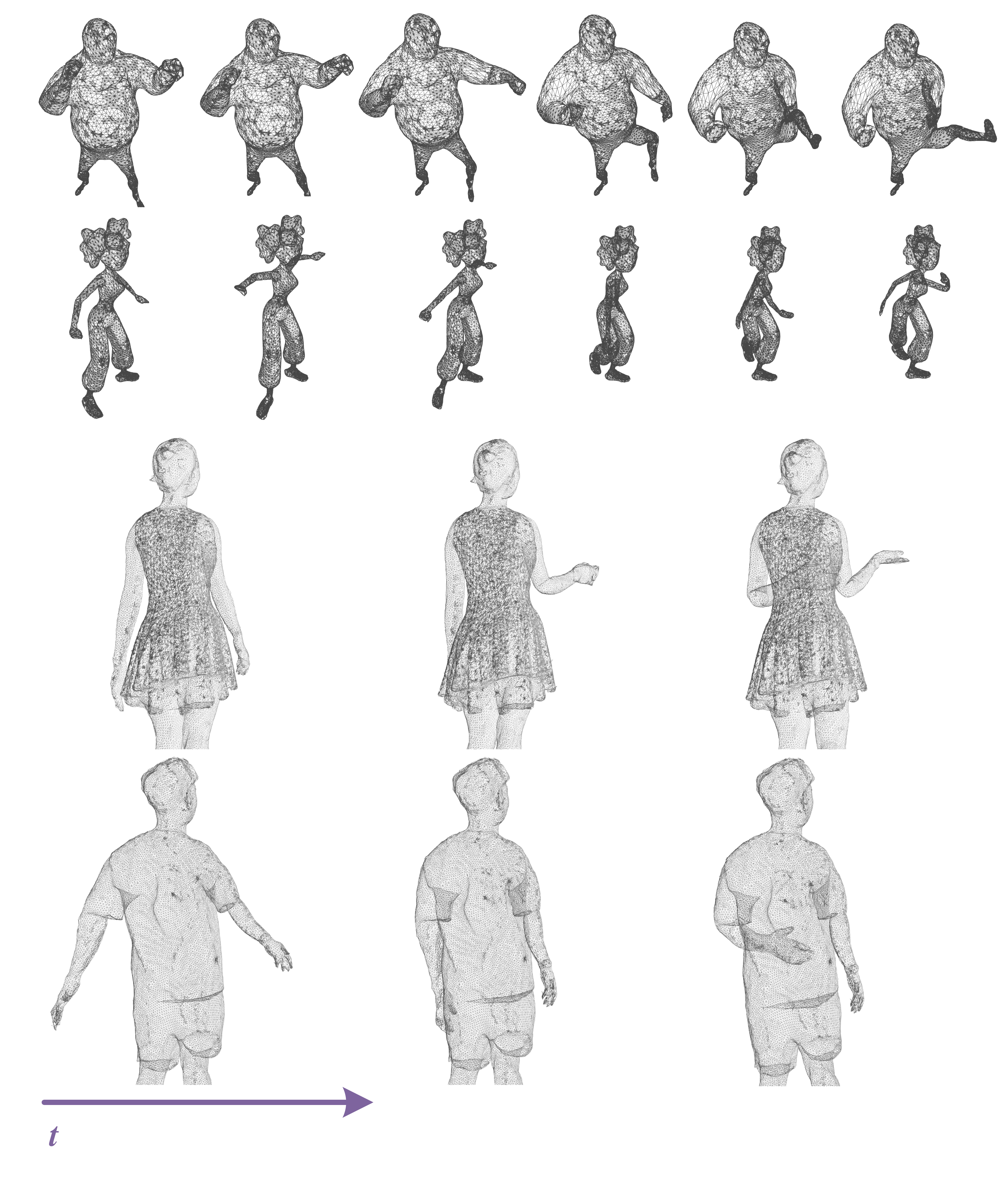}
    \vspace{-0.9cm}
    \caption{The topology of the mesh sequence.}
    \vspace{-0.6cm}
    \label{fig:Topology-consistent mesh}
\end{figure}

\begin{figure}[h]
    \centering
    \includegraphics[width=\linewidth]{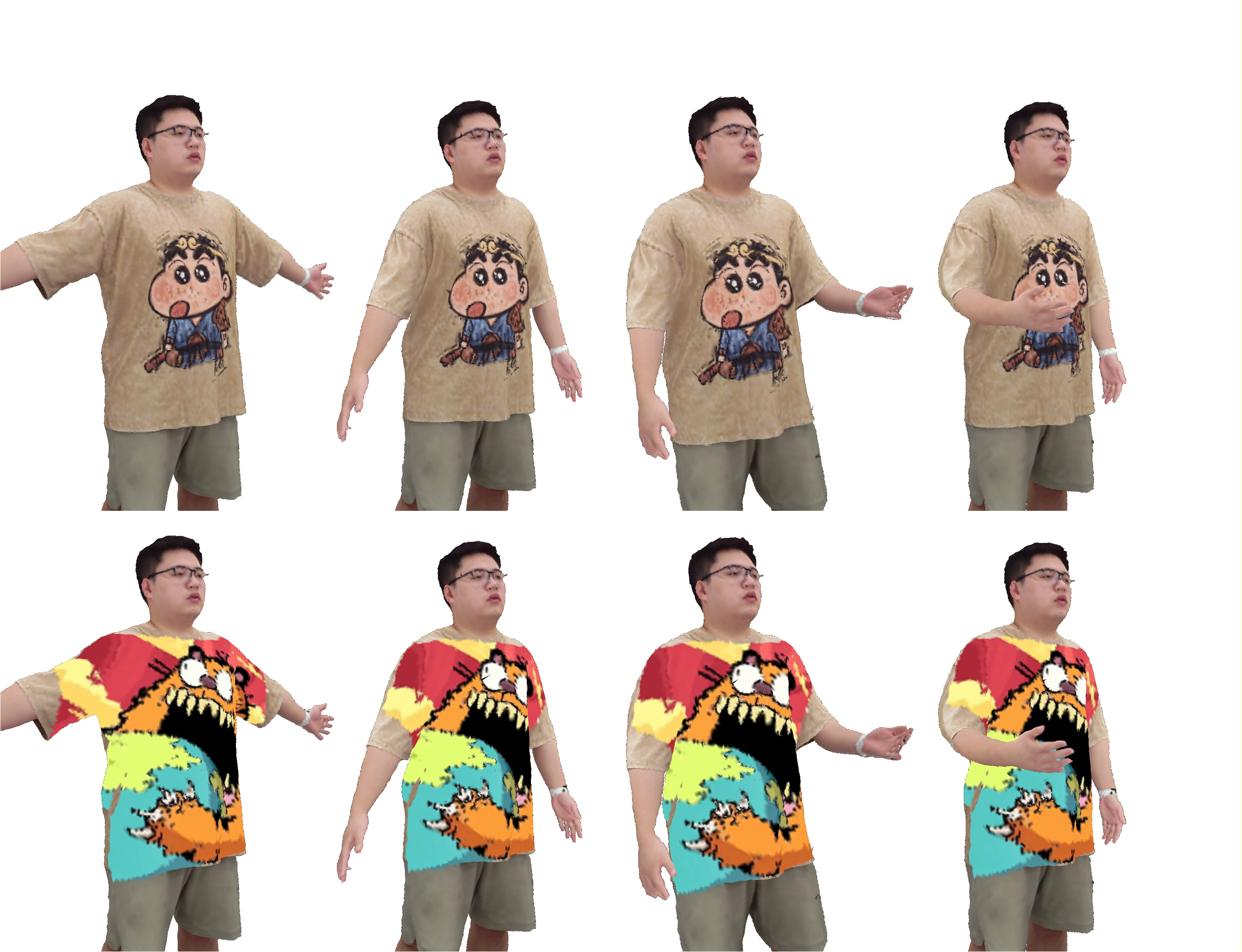}
    \caption{Model Editing Result.}
    \label{fig:Model_Editing}
\end{figure}

\section{Topology-consistent model sequence} \label{sec:sup_D}

A key advantage of our method is its ability to reconstruct a sequence of topology-consistent 3D models from multi-view videos of dynamic objects. This property is essential for downstream tasks such as geometric analysis, motion processing, and physical simulation. As shown in \cref{fig:Topology-consistent mesh}, the reconstructed sequence maintains identical vertex counts and connectivity across all frames, ensuring consistent mesh topology throughout the motion.

\section{Efficient Editing of Dynamic Sequences via Consistent Topology} \label{sec:sup_F}

Our method ultimately produces topology-consistent Gaussian and mesh sequences of the target object. The globally consistent topology allows reliable 3D keypoint tracking and significantly simplifies the editing of dynamic sequences. Specifically, since all frames share the same topology, editing a single frame and propagating the changes across the sequence is sufficient to achieve global adjustments to the dynamic model. The results of model editing are shown in \cref{fig:Model_Editing}.


\end{document}